\documentclass[%
 aip,
 jcp,
 amsmath,amssymb,
 reprint,
 floatfix, 
]{revtex4-1}

\usepackage{graphicx}
\usepackage[usenames,dvipsnames]{color}
\usepackage{algorithm}
\usepackage{algpseudocode}
\usepackage{xcolor}
\usepackage{bm}
\usepackage[normalem]{ulem}
\usepackage{multirow}
\usepackage{upgreek} 
\usepackage[pdftex]{}
\usepackage[colorlinks,allcolors=black,pdftex]{hyperref}
\usepackage{xr}

\newcommand{\equalcontrib}{These authors contributed equally to this work.}

\begin{document}

\preprint{AIP/123-QED}
\title{How Thermostats Influence Dynamics Across Time Scales: A Systematic Study from Fast Motions to Slow Transitions}

\author{Frederick Heinz}
\thanks{\equalcontrib}
\affiliation{ 
Freie Universität Berlin, Department of Biology, Chemistry and Pharmacy, Arnimallee 22, 14195 Berlin}

\author{Sascha Jähnigen}
\thanks{\equalcontrib}
\affiliation{ 
Freie Universität Berlin, Department of Biology, Chemistry and Pharmacy, Arnimallee 22, 14195 Berlin}

\author{Joana-Lysiane Schäfer}
\thanks{\equalcontrib}
\affiliation{ 
Freie Universität Berlin, Department of Biology, Chemistry and Pharmacy, Arnimallee 22, 14195 Berlin}

\author{Bettina G.~Keller}%
\email{bettina.keller@fu-berlin.de}
\affiliation{ 
Freie Universität Berlin, Department of Biology, Chemistry and Pharmacy, Arnimallee 22, 14195 Berlin}

\date{\today}

\begin{abstract}
\section*{Abstract}
Reliable dynamical properties from molecular dynamics simulations require careful control of thermostatting artifacts.
We systematically assess how NVE, deterministic thermostats, velocity-rescale dynamics, and stochastic Langevin-type thermostats affect time-correlation functions across liquids of varying complexity. 
The analysis spans vibrational spectra, velocity and pressure autocorrelations, diffusion coefficients, shear viscosities, and Markov state models. 
Deterministic thermostats and velocity-rescale dynamics closely reproduce NVE reference data over all observables. In contrast, strongly coupled stochastic thermostats ($\tau_T < 1\, \mathrm{ps}$) systematically distort dynamical properties. 
By constrast, moderate stochastic coupling ($\tau_T \approx 1\, \mathrm{ps}$) restores near-NVE behavior while maintaining canonical sampling.
Our results provide practical guidelines for selecting thermostat schemes when accurate dynamical properties or Markov models are required.
\end{abstract}

\maketitle

\section{Introduction}

Transport properties obtained from molecular dynamics (MD) simulations via time correlation functions have gained considerable interest in recent years. 
Time-correlation functions provide a direct link between microscopic dynamics sampled by the MD simulation and macroscopic kinetic properties.
This approach is grounded in the Onsager Regression Hypothesis \cite{OnsagerI, OnsagerII, Kubo1966}, which states that the decay of spontaneous equilibrium fluctuations mirrors the relaxation behavior of a macroscopic system after a small perturbation.
Time-correlation functions can be used to model a vast range of phenomena. 
Examples include 
vibrational spectra (infrared and circular dichroism)\cite{Gaigeot2021,Abbate1998,Ivanov2013,Ditler2022,Thomas2013,Jaehnigen2025}, 
structure factors in liquids and solids \cite{Fransson, moustafa2018effects},
thermal conductivity \cite{knoop2023ab, Olarte-Plata02012022},
NMR relaxation times \cite{lipari1982model, hoffmann2020predicting},
diffusion constants \cite{Heinz24},
viscosities \cite{hess2002determining, maginn2019best}
slip lengths and shear stress in non-equilibrium settings \cite{Maffioli} and 
conformational transitions rates in Markov models \cite{keller2010comparing, prinz2011markov}.
From ultrafast vibrational motions to slow conformational changes, the decay times of these correlation functions can vary by orders of magnitude, which poses considerable challenges for sampling and analysis in MD simulations.
The effect of a thermostat on time-correlation functions and the associated dynamic properties of a system is a widely discussed topic. 
It often forms part of the discussion of a newly introduced integration algorithm \cite{Nose-Hover, bussi2007canonical, hunenberger2005thermostat}.
In particular, tightly coupled stochastic thermostats are known to distort the diffusion constant and viscosity estimates of water and other simple liquids \cite{bussi2008stochastic,Junghans2007,Ruiz-Franco2018,Basconi2013,Thermostats22}.
To mitigate these artifacts, several correction schemes have been proposed \cite{hicks2021removing}, and a number of advanced Langevin integrators have been designed to improve the accuracy of dynamical observables \cite{bussi2007canonical, sivak2014time}.

Depending on the specific observable and the practices of the respective community, different modi operandi have emerged.
For the calculation of vibrational spectra, that uses \textit{ab initio} molecular dynamics (AIMD), it is common practice to accomplish "proper" canonical sampling through averaging over a series of statistically independent NVE trajectories that have been  sampled from a parent NVT trajectory using a thermostat \cite{Ivanov2013,Gaigeot2021,Jähnigen2021,Ditler2022}.
By contrast, when simulating slow processes in soft-matter systems, the standard is to use deterministic thermostats. or the velocity rescale-thermostat \cite{bussi2007canonical}. 
But also stochastic thermostats with a sufficiently weak thermostat coupling constant are frequently employed \cite{leimkuhler2013robust}.
A second major challenge in accurately estimating time-correlation functions arises from slow conformational transitions. 
In the absence of enhanced sampling techniques, the resulting correlation functions are heavily influenced by the initial distribution of conformations and may be significantly biased. 
To overcome this, dynamical reweighting techniques \cite{keller2024dynamical} can be employed to recover accurate time-correlation functions from biased simulations.
In particular, Girsanov reweighting \cite{donati2018girsanov, Luca}, in which the reweighting factors are derived from stochastic path measure of the simulated trajectory, can be used as a variance reduction methods for calculating transport properties of stochastic dynamics \cite{gastaldello2025dynamical}.
Recent applications include kinetic force-field optimization \cite{bolhuis2023optimizing}, catalytic cycle acceleration \cite{bolhuis2025optimal}, and machine-learning of slow collective variables \cite{shmilovich2023girsanov}.
However, such reweighting methods require stochastic thermostats, which are typically implemented via direct numerical integration of Langevin dynamics  \cite{Luca, keller2024dynamical}, which unavoidably perturb the underlying transport properties to some extent.
Compared to deterministic thermostats, stochastic thermostats offer several additional advantages: they inherently preserve ergodicity and typically ensure robust sampling of equilibrium properties, even under variations in simulation parameters such as thermostat coupling strength or time step \cite{leimkuhler2013robust, fass2018quantifying, finkelstein2021bringing, zhang2019unified}. 
These benefits make stochastic thermostats highly attractive—even for simulations aimed at extracting transport properties.
Yet, a systematic and comprehensive assessment of the magnitude of thermostat-induced distortions across different transport properties and molecular systems is still lacking.

The aim of this study is to benchmark how NVE simulations, as well as commonly used deterministic and stochastic thermostats, affect the accuracy of dynamic properties in molecular dynamics. 
We focus on thermostat schemes readily available in standard MD packages and probe whether deviations arise from interference between the thermostat coupling time and the intrinsic timescale of the observable.
To this end, we analyze time-correlation functions spanning a wide range of decay times—from the sub-picosecond velocity autocorrelation function to the nanosecond pressure autocorrelation function of viscous liquids.
We further include dynamical observables that involve more complex transformations than simple time integrals, such as the vibrational density of states and slow processes obtained from Markov state models.

\section{Theory}

\subsection{Molecular dynamics}
Consider a system of $N$ atoms, where $\mathbf{q}_i = (q_{ix}, q_{iy}, q_{iz})$ denotes the position vector of the $i$-th particle in the three-dimensional Cartesian space. 
The equations of motion for an isolated system are Newton's laws of motion,
\begin{eqnarray}
    \dot{\mathbf{q}}_i &=& \frac{\mathbf{p}_i}{m_i} \cr
    \dot{\mathbf{p}}_i &=& - \nabla_i V(\mathbf{q}) \, ,
\label{eq:Newton}    
\end{eqnarray}
where we used the dot notation for the time derivative $\dot{\mathbf{q}}_i = \partial \mathbf{q}_i / \partial t$.
In the first line, $\dot{\mathbf{q}}_i$ is the velocity of the $i$-th particle, $\mathbf{p}_i$ its momentum and $m_i$ its mass. 
In the second line, 
$\dot{\mathbf{p}}_i$ denotes the force acting on particle $i$,
due to the potential energy function $V(\mathbf{q})$, whose gradient 
$\nabla_{i} = \partial/\partial{\mathbf{q}_{i}}$ is taken with respect to the coordinates of the $i$-th particle.
For a system in contact with a heat bath at temperature $T$, 
the equations of motion can be modeled by underdamped Langevin dynamics,
\begin{align}
    \dot{\mathbf{q}}_i &= \frac{\mathbf{p}_i}{m_i} \,, \cr
    \dot{\mathbf{p}}_i &= - \nabla_i V(\mathbf{q}) - \xi \mathbf{p}_i 
    + \sqrt{2\xi k_{\mathrm{B}} T m_i}\,\boldsymbol{\upeta}_i(t)\,,
\label{eq:underdampedLangevin}
\end{align}
where $\xi$ is the collision or friction rate (in units of time$^{-1}$), 
$T$ the temperature, and $k_{\mathrm{B}}$ the Boltzmann constant.
The term $\boldsymbol{\upeta}_i(t) = \bigl(\upeta_{ix}(t), \upeta_{iy}(t), \upeta_{iz}(t)\bigr)$ 
denotes Gaussian white noise with
$\bigl\langle \upeta_{i\alpha}(t) \bigr\rangle = 0$
$\bigl\langle \upeta_{i\alpha}(t)\,\upeta_{j\beta}(t') \bigr\rangle = \delta_{ij}\,\delta_{\alpha\beta}\,\delta(t-t')$,
for $\alpha,\beta \in \{x,y,z\}$, 
where $\delta_{ij}$ and $\delta_{\alpha\beta}$ are Kronecker deltas and $\delta(t-t')$ 
is the Dirac delta function.

In the second line of eq.~\ref{eq:underdampedLangevin}, the total force acting on particle $i$ is composed of the force due to the potential energy (first term), the friction force (second term), and a compensating random force (third term).
The relative magnitudes of  friction and random force are derived via the dissipation-fluctuation theorem.\cite{Kubo1966}
Their relation ensures that the average temperature of the system is proportional to its average 
kinetic energy
\begin{eqnarray}
    \langle T\rangle &=& \frac{2}{3Nk_B} \, \langle E_{\mathrm{kin}} \rangle
\label{eq:average_T}    
\end{eqnarray}
where the kinetic energy is
$E_{\mathrm{kin}} =\sum_{i=1}^N  \frac{\mathbf{p}_i^{\top}(t)\mathbf{p}_i(t)}{2m_i}$, and $\langle ... \rangle$ denotes an ensemble average.

There are three points to note here:
(i) Eq.~\ref{eq:Newton} samples the microcanonical ensemble ($NVE$), whereas
eq.~\ref{eq:underdampedLangevin} samples the canonical ensemble ($NVT$).
(ii) By setting the frequency of collisions with the heat bath, $\xi$, to zero in eq.~\ref{eq:underdampedLangevin}, the equation of motion for an isolated system (eq.~\ref{eq:Newton}) is recovered.
(iii) While the fluctuation-dissipation theorem defines the magnitude of the random force, its direction is uniformly distributed, independent of both the potential-derived force and the friction force.

\subsection{MD integrators}
Various MD integrators\cite{matthews2015molecular, frenkel2023understanding, tuckerman2010statistical} have been designed to numerically solve the equations of motion for both the NVE and NVT ensembles (eq.~\ref{eq:Newton} and \ref{eq:underdampedLangevin}).

\paragraph*{NVE ensemble (NVE).}
We use the Leap Frog and the Velocity Verlet algorithm \cite{verlet1967computer, matthews2015molecular, frenkel2023understanding, tuckerman2010statistical} to update the positions and momenta according to eq.~\ref{eq:Newton}.
Both algorithms yield the same position trajectory.
Velocity Verlet updates positions and velocities at the same time step, requiring two force evaluations per step, while Leap-Frog updates velocities at half-time steps, requiring only one force evaluation but needing interpolation to obtain velocities at full time steps.
They do not apply random numbers. 
\paragraph*{NVT ensemble / stochastic dynamics (SD).}
We use GROMACS Stochastic Dynamics\cite{goga2012efficient} and the OBABO splitting method \cite{bussi2007accurate, bussi2007canonical, matthews2015molecular} to update the positions and momenta according to eq.~\ref{eq:underdampedLangevin}.
GROMACS Stochastic Dynamics is equivalent to the BAOA splitting method \cite{Kieninger2022} and applies a single random number for each integration step and thermostated degree of freedom.
The OBABO splitting method applies two random numbers for each integration step and thermostated degree of freedom.
In practice, the OBABO splitting method is implemented as the Velocity Rescale thermostat\cite{bussi2007accurate, bussi2007canonical, bussi2008stochastic}, where setting the number of thermostated degrees of freedom to one yields the OBABO scheme (see Sec.~\ref{sec:Theory_thermostates}).
%

%
To compute the random force in eq.~\ref{eq:underdampedLangevin}, SD requires the generation of random numbers at each time step for all dimensions that are coupled to the heat bath \cite{goga2012efficient, fass2018quantifying, gronbech2024definition, matthews2015molecular}.
This sequence of random numbers forms a dynamic process in itself which interacts with the dynamics of the system, potentially distorting the correlation functions. 
As a result, dynamic and transport properties derived from these functions may be inaccurate \cite{basconi2013effects}.
%

\subsection{MD thermostats}
\label{sec:Theory_thermostates}
We here take the viewpoint that thermostat algorithms are MD integrators for systems with thermal coupling that minimize or eliminate the use of random numbers. 
Instead of using friction and random forces to model thermal coupling, these algorithms apply alternative force terms, leading to modified equations of motion.
The following sections list the thermostats and their equations of motion that we use in this study.

\paragraph{Berendsen thermostat.}\cite{Berendsen1984} (BE) 
The particle momenta are scaled such that the instantaneous temperature $T(t)$ relaxes to the target temperature $T$
\begin{eqnarray}
   \dot{\mathbf{q}}_i &=& \frac{\mathbf{p}_i}{m_i}\cr
    \dot{\mathbf{p}}_i &=& - \nabla_i V(\mathbf{q}) - \xi \left( \frac{T}{T(t)} -1  \right) \,  \mathbf{p}_i \, .
\label{eq:Berendsen}
\end{eqnarray}
where the instantaneous temperature of the system can be defined via its kinetic energy at time $t$: 
$E_{\mathrm{kin}}(t)  = \frac{3N}{2} k_B\, T(t)$. 
No random forces are applied.

%
%
\paragraph{Velocity–rescale thermostat}\cite{bussi2007accurate, bussi2007canonical, bussi2008stochastic}
(VR; also referred to as the Bussi–Donadio–Parrinello thermostat).
In this modification of the Berendsen scheme, an additional stochastic term is introduced to ensure sampling of the canonical ensemble.
The equations of motion can be written as
\begin{align}
    \dot{\mathbf{q}}_i &= \frac{\mathbf{p}_i}{m_i}\,, \cr
    \dot{\mathbf{p}}_i &= - \nabla_i V(\mathbf{q})
    + \xi \left[ \left( 1 - \frac{1}{N_f} \right)\frac{T}{T(t)} - 1 \right] \mathbf{p}_i  \cr
    &\quad\; + \sqrt{ \frac{2\xi}{N_f}\frac{T}{T(t)} }\, \mathbf{p}_i \, \upeta(t)\,,
\label{eq:velocityRescale}
\end{align}
where $N_f$ is the number of unconstrained degrees of freedom coupled to the heat bath.
The term $\eta(t)$ is a \emph{single global Gaussian white-noise process} with $\langle \upeta(t) \rangle = 0$ and $\langle \upeta(t)\,\upeta(t') \rangle = \delta(t-t')$
and the same noise realization is applied to all thermally coupled degrees of freedom.
Thus, the velocity–rescale thermostat draws \emph{one} random number per integration step and rescales all relevant momenta consistently.
For $N_f = 1$, noting that $p_{i\alpha}^2/m_i = k_{\mathrm{B}}T(t)$ (for $\alpha = x,y,z$), 
eq.~\ref{eq:velocityRescale} reduces to the underdamped Langevin equation 
eq.~\ref{eq:underdampedLangevin} with $\eta(t)$ as the random force.

%
\paragraph{Nos\'e-Hoover thermostat}\cite{Nose-Hover} (NH).
The phase space is extended by a 
pseudo-particle with variable $\upeta \in \mathbb{R}$ and coupling parameter $Q$. The friction variable $p_{\upeta}$ is driven by the difference between the instantaneous value of the kinetic energy and its canonical average.\cite{tuckerman2010statistical}
It does not apply random numbers.

\begin{eqnarray}
    \dot{\mathbf{q}}_i  &=& \frac{\mathbf{p}_i}{m_i}\cr
    \dot{\mathbf{p}}_i  &=& - \nabla_i V(\mathbf{q}) - \frac{p_{\upeta}}{Q} \mathbf{p}_i  \cr
    \dot{\upeta}          &=& \frac{p_{\upeta}}{Q} \cr
    \dot{p}_{\upeta}      &=& \sum_{i=1}^N m_i \dot{\mathbf{q}}_i^2 - 3N_fk_BT\, .
\label{eq:NoseHoover}
\end{eqnarray}
Although lines 3 and 4 resemble the definitions of pseudo-particle momentum and total force, they do not represent them. 
Caution is needed to avoid this misconception.
$\upeta$ is dimensionless. 
Consequently, $\dot{\upeta}$ has units of time$^{-1}$. 
From the r.h.s.~of the fourth line, we can see that $\dot{p}_{\upeta}$ has units of energy, and thus $p_{\upeta}$ has units of energy$\times$time. 
From this we can conclude that the coupling parameter $Q$ has units of energy$\times$time$^2$.
It can reformulated in terms of a collision frequency $\xi$ and a target temperature $T$ as\cite{Nose-Hover,tuckerman2010statistical}
\begin{eqnarray}
    Q &=& \frac{N_f k_B}{4\pi^2} \frac{T}{\xi^2} \, .
\label{eq:NH_couplingParameter}    
\end{eqnarray}
The factor $p_{\upeta}/Q$ in line 3 couples the dynamics of the pseudo-particle to the system. 
In contrast to a collision frequency, it can be positive and negative, and thus this second term in line 3 can drain energy from the system or feed energy into it. 
Line 4 couples the dynamics of the system to the pseudo particle: $\dot{p}_{\upeta}$ is twice the deviation between current kinetic energy of the system and the canonical average of the kinetic energy at the target temperature. 
Due to this linear coupling, MD simulations with NH thermostat may exhibit temperature oscillations, but the thermostat does sample the correct canonical distribution.

%
%
\paragraph{Temperature decay and coupling time.} 
For underdamped Langevin dynamics, the temperature decay $T(t)$ of a system initialized at at $T(t=0)$ towards the temperature of the heat bath $T$ is
\begin{eqnarray}
    \langle \dot{T}(t)  \rangle
    &=&  2\xi \left( T - \left\langle T(t) \right\rangle \, \right) \, ,
\label{eq:Tcoupling_underampedLangevin}    
\end{eqnarray}
where the friction coefficient $\xi$ determines how tightly the system is coupled to the heat bath.
In GROMACS\cite{GROMACS_REFERENCE}, the Langevin integrator GSD is controlled by the inverse of $\xi$, the coupling time $\tau_T^{\mathrm{GSD}} = \xi^{-1}$. 
VR and B thermostats are designed to yield the same temperature decay as in eq.~\ref{eq:Tcoupling_underampedLangevin}.
However, by convention, the coupling strength is controlled by $\tau_T^{\mathrm{VR}} = \tau_T^{\mathrm{B}} = (2\xi)^{-1}$.
Thus, to achieve the same coupling strength in GSD, VR and B, the coupling times need to be chosen as: 
$\tau_T^{\mathrm{VR}} = \tau_T^{\mathrm{B}} = \frac{1}{2}\tau_T^{\mathrm{GSD}}$.

\begin{figure}[h]
    \centering
    \includegraphics[width=\columnwidth]{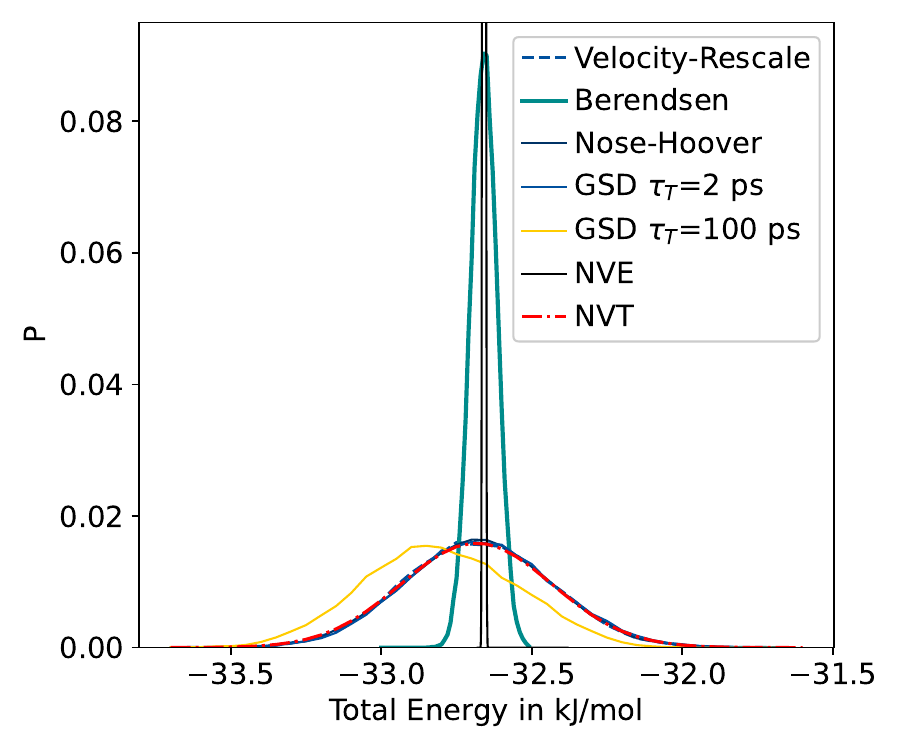}\\
    \caption{Energy distribution of TIP3P water in NVE and NVT ensemble with different thermostats with a simulation time of 60~ns. The coupling time $\tau_T$ is 1 ps for Berendsen and velocity rescale and 2 ps for GROMACS stochastic dynamics GSD. The chain length for Nose-Hoover is 1 with a coupling time of 4 ps. NVT is calculated using a normal distribution and eq.~\ref{eq:sigma}  with a Cv of 4.14~kJ/kgK\cite{Mao2012}}
    \label{fig:energyDistribution}
\end{figure}

%
\paragraph{Energy fluctuations and ensembles.} 
LF and velocity-verlet\cite{VV} algorithm simulate an isolated system in the $NVE$ ensemble, where the total energy remains constant and its distribution approximates a delta function around the mean energy $\langle E \rangle$. 
In contrast, GSD, VR, and NH sample the $NVT$ ensemble \cite{GROMACS}\cite{Nose-Hover}\cite{bussi2008stochastic}, where the total energy fluctuates around its mean with a variance related to the heat capacity $C_V$\cite{Cv}.
\begin{eqnarray}
    \sigma_E^2 = \langle E^2 \rangle - \langle E \rangle^2 = k_BT^2C_V \, .
    \label{eq:sigma}
\end{eqnarray}
B maintains the average temperature (eq.~\ref{eq:average_T}) but does not sample a conventional thermodynamic ensemble \cite{Berendsen1984, Hummer}, resulting in an energy distribution that lies between the $NVE$ and $NVT$ limits.
Fig.~\ref{fig:energyDistribution} demonstrates this for a water simulation.

\subsection{Correlation functions}
Observables $f, g$ are functions that assign a real number to the state of the system $(\mathbf{q}, \mathbf{p})$. 
As the system evolves over time, these observables also change accordingly.
The short-hand notation $f(\mathbf{q}(t), \mathbf{p}(t))=f(t)$ is used, with an analogous notation for $g$.
The time-dependent correlation function quantifies the dynamical relationship between two observables
\begin{eqnarray}
    \langle f(0)g(\tau) \rangle &=& \lim_{T \to \infty}\frac{1}{T}\int_0^T f(t)g(t+\tau) \, \mathrm{d}t \, ,
\label{eq:correlationFct}    
\end{eqnarray}
where $\langle \dots \rangle$ again denotes an ensemble average, and the equality holds if the process is ergodic.
If both observables are the same, by setting $g=f$ in eq.~\ref{eq:correlationFct}, one obtains the auto-correlation function (ACF) of $f$. 
It describes how the observable $f$ retains its value over time.
In linear response theory, time correlation functions are directly related to the (time-lagged) response of a system to a weak perturbation and thus establish a link between theory and experiment.\cite{Berne1971}
In MD simulations, correlation functions are used to construct Markov state models\cite{swope2004describing, buchete2008coarse, keller2010comparing, prinz2011markov}, capturing slow dynamics beyond linear response, while autocorrelation functions provide transport coefficients through Green–Kubo relations and determine spectral density functions \cite{Kubo1966, frenkel2023understanding, tuckerman2010statistical}.

\subsection{Diffusion constant}
Consider an unevenly distributed substance dissolved in a fluid, where  $c(x,y,z,t)$ is concentration of the substance at point $(x,y,z)$ and time $t$. 
Moreover, assume that the particles of this substance evolve according to diffusive dynamics.
Fick's second law of diffusion states that the concentration changes over time as
\begin{eqnarray}
    \dot{c} &=& D \,\nabla^2 c\, ,    
\end{eqnarray}
where $\nabla^2$ is the Laplacian operator defined as $\nabla^2 = \frac{\partial^2}{\partial x^2} + \frac{\partial^2}{\partial y^2} + \frac{\partial^2}{\partial z^2}$.
$D$ is the diffusion coefficient, which has units of area per unit time.
It can be calculated from the velocity fluctuations of a single particle via the following Green-Kubo relation\cite{Kubo1966},
\begin{equation}
    D= \frac{1}{3}\int_{0}^{\infty} \langle \mathbf{v}^{\top}_{\mathrm{COM}}(0)\, \mathbf{v}_{\mathrm{COM}}(\tau) \rangle  \,\mathrm{d}\tau \, ,
\label{eq:GK-self-diffusion}
\end{equation}
where $\mathbf{v}_{\mathrm{COM}}(t)$ is the velocity vector of the particle's center of mass and $\langle \mathbf{v}_{\mathrm{COM}}(0)\, \mathbf{v}_{\mathrm{COM}}(\tau) \rangle$ is its ACF (analogous to eq.~\ref{eq:correlationFct}).
The factor $1/3$ accounts for the fact that the particle diffuses in three dimensions.
The concept of a diffusion coefficient can be generalized to the diffusion of a particle in a fluid of the same particles. 
In this case, the concentration gradient is zero, nonetheless the self-diffusion coefficient $D$ can be defined via eq.~\ref{eq:GK-self-diffusion}.
\subsection{Shear viscosity}
Consider a fluid confined between two infinite, parallel planes separated by distance $h$ along the $z$-axis. 
The upper plane moves at constant velocity $u_x$ in $x$-direction and exerts a constant shear force per unit area $F_x/A$, which is also called the shear stress.
As a result, the uppermost layer of the fluid moves in direction $x$ with the same velocity $u_x(z=h)$ as the plate, whereas the lowermost layer remains at rest $u_x(z=0)=0$. 
In Newtonian fluids, the resulting velocity gradient is constant and is related to the shear stress by
\begin{eqnarray}
    \frac{F_x}{A} &=& \eta \frac{\partial u_x}{\partial z} \, .  
\end{eqnarray}
$\eta$ is the dynamic viscosity of the fluid with units of mass per distance per time squared.
The dynamic viscosity can be calculated from the fluctuations of the pressure tensor via the following Green-Kubo relation
\cite{Kubo1966, Hansen2013, frenkel2023understanding}
\begin{equation}
    \eta= \frac{1}{Vk_BT} \int_0^{\infty} \langle P^{\alpha\beta}(0){P^{\alpha\beta}(\tau)}\rangle \, \mathrm{d}\tau \, ,
    \label{eqn:GK_viscosity}
\end{equation}
where $\alpha, \beta = x,y,z$ and $V$ is the volume of the system, $P^{\alpha\beta}$ an off diagonal element of the system pressure tensor, and 
$\langle P^{\alpha\beta}(0){P^{\alpha\beta}(\tau)}\rangle$ is its ACF. The system pressure tensor is symmetric. To obtain the correct viscosity, the results need to be averaged over all three off diagonal elements $P^{xy}$, $P^{yz}$ and $P^{xz}$.

\subsection{Vibrational density of states }
\label{sec:vDOS}
For \textit{time-periodic} perturbations of the form $F_0\cos(\omega t)$, where $F_0$ is the field amplitude, the system's response gives rise to an energy absorption spectrum, which does not correspond to a transport property.
Instead, it describes the rate at which energy is dissipated with a frequency $\omega$ due to induced transitions between the quantum eigenstates of the system. 
In the context of time-dependent perturbation theory, the corresponding average transition rate is defined via Fermi's Golden Rule,\cite{tuckerman2010statistical,McQuarrie2000}
\begin{equation}
    R(\omega) = \frac{2\pi}{\hbar^2} \left|F_0\right|^2  \sum_{i,f} \rho_i |\langle f|\hat{\mathcal{V}}| i \rangle |^2 \delta(\omega_{if} -\omega),
    \label{eq:fermi_GR}
\end{equation}
where $\hbar$ is the reduced Planck number and $\rho_i$ is the probability of finding the system in the initial state. The labels $i$ and $f$ denote the initial and final eigenstate, respectively, with $\omega_{fi}$ being the angular transition frequency between them, $\omega_{fi}= \frac{1}{\hbar}E_{fi} = \frac{1}{\hbar}(E_f - E_i)$.
The delta function ensures that transitions occur only when the energy difference matches the energy of the perturbation.
$\langle f|\hat{\mathcal{V}}| i \rangle$ is the matrix element of the perturbation between the initial and final states with the Hermitian operator $\hat{\mathcal{V}}$ defining the response of the system.

If the time scale of the perturbation is within the range of the nuclear dynamics and the states $i$ and $f$ correspond to nuclear eigenstates, eq.~\ref{eq:fermi_GR} can be estimated by means of MD simulations.
Writing the delta function as an integral, $\delta(\omega) = \frac{1} {2\pi}\int_{-\infty}^{\infty} \exp(-i \omega t) \mathrm{d} t$, eq.~\ref{eq:fermi_GR} can be reformulated in the Heisenberg picture as the Fourier transform of the quantum time correlation function (TCF),\cite{gordon1965,Berne1971,tuckerman2010statistical}
\begin{equation}
    R(\omega) = \frac{\left|F_0\right|^2}{\hbar^2}   \int_{-\infty}^{\infty} \mathrm{e}^{-\mathrm{i} \omega t} \langle\hat{\mathcal{V}}(0)\hat{\mathcal{V}}(t)\rangle \mathrm{d} t,
    \label{eq:fermi_GR_IP}
\end{equation}
where the explicit sum over initial/final states is now described by means of a quantum time-correlation function of the response operator $\hat{\mathcal{V}}$.
From this, the absorption spectrum can be derived as the net energy flow per unit time, determined by the detailed balance between absorption and emission processes in a canonical distribution: $Q(\omega) = \hbar \omega  \left(1 - \mathrm{e}^{-\beta\hbar\omega}\right) R(\omega)$. 

In classical MD simulations, a convenient approach to estimate eq.~\ref{eq:fermi_GR_IP} is to derive its classical limit, where the quantum operator can eventually be replaced by classical phase space functions.\cite{Kubo1957,Ramirez2004}
Hence, the absorption spectrum becomes,\cite{tuckerman2010statistical}
\begin{align}
    Q(\omega) &=  \frac{\left|F_0\right|^2  }{k_B T} \omega^2\int_{-\infty}^{\infty} \mathrm{e}^{-\mathrm{i} \omega t} \langle\mathcal{V}(0)\mathcal{V}(t)\rangle \mathrm{d} t\cr
    &= \frac{\left|F_0\right|^2 }{k_B T}   \int_{-\infty}^{\infty} \mathrm{e}^{-\mathrm{i} \omega t} \langle\dot{\mathcal{V}}(0)\dot{\mathcal{V}}t)\rangle \mathrm{d} t.
    \label{eq:Q_w}
\end{align}
The second line follows from the properties of the Fourier transform and the stationarity of the TCF, which is now a classical one. Eq.~\ref{eq:Q_w} is an expression of the Wiener-Khinchtine theorem.\cite{McQuarrie2000}

Vibrational spectroscopies, including techniques such as inelastic neutron scattering, probe the vibrational eigenstates of a system. The system response is linked to the vibrational density of states (vDOS), denoted as $g(\omega)$, which describes the distribution of vibrational states as a function of frequency,
\begin{equation}
    g(\omega) = \frac{1}{3N} \sum_{i=1}^{3N} \frac{m_i}{k_B T} \int_{-\infty}^{\infty} \langle \dot{q}_i(0) \dot{q}_i(t) \rangle \text{e}^{-\mathrm{i}\omega t} dt,
\label{eq:vdos}
\end{equation}
where  $\dot{\bm{q}}(t) = \partial{\bm{q}}/\partial t(t) = (\dot{q}_1(t), ..., \dot{q}_{3N}(t))$ denotes the nuclear velocities at time $t$,
$m_i$ is the particle mass, and the average has been taken over all $3N$ degrees of freedom.
Eq.~\ref{eq:vdos} can be understood as the system's response to a time-dependent external driving force $F(t)$ (\textit{e.g.}, exerted by an electric field, neutrons, \textit{etc.}), which induces a displacement, so that $\mathcal{V}(t) = \bm{q}(t)$.
Eq.~\ref{eq:vdos} is obtained from eq.~\ref{eq:Q_w} by normalizing with the absorbed power amplitude, $F_0^2/(\omega m_i)$, and subsequently dividing by $\omega$, in order to obtain a quantity expressed as a spectral density in frequency.

In an alternative procedure, the spectral density can be derived from the susceptibility function derived from the (quantum) linear response function, which leads to the same result.\cite{tuckerman2010statistical,Berne1970}
The correspondence between the classical expression and the quantum vDOS is valid only within the harmonic approximation. However, anharmonic effects are correctly described in terms of the (shifted) position of the fundamental frequencies.\cite{goncalves1992,dickey1969}

It should be noted that vDOS intensities cannot be directly compared to intensities from infrared absorption, Raman scattering, or vibrational optical activity, nor to the corresponding spectral line shapes.\cite{Bowles2023,Thomas2013,Ivanov2013}

\subsection{Markov state models}
Markov state models (MSMs) \cite{swope2004describing, buchete2008coarse, prinz2011markov, lemke2016density} provide a framework for modeling the microscopic dynamics of a system, such as the conformational transitions of a single molecule immersed in a solvent. To construct an MSM, the system’s state space $\Gamma$ is projected on a low-dimensional collective variable (CV) space: $\mathbf{x}: \Gamma \rightarrow \mathbb{R}^m$, where $x$ is an $m$-dimensional CV vector with $m \ll 3N$.
It is then assumed that the dynamics in this reduced space can be effectively described by a Markovian process.

For example, consider that the effective dynamics in the CV space follows overdamped Langevin dynamics. This process is governed by an associated Fokker-Planck equation, which describes the time evolution of the probability density $\rho(\mathbf{x},t)$ in the CV space
\begin{subequations}
\begin{eqnarray}
    \dot{\rho}(\mathbf{x},t) &=& \mathcal{L}\rho(\mathbf{x},t)  \label{eq:MSM_FP}\\
    \rho(\mathbf{x},t+\tau) &=& \text{e}^{\mathcal{L} \tau} \rho(\mathbf{x},t) \label{eq:MSM_propagator}
\end{eqnarray}
\end{subequations}
where $\mathcal{L}$ is the Fokker-Planck operator and $\mathcal{P}(\tau) =  \text{e}^{\mathcal{L} \tau}$ is the corresponding propagator with lag time $\tau$.
$\mathcal{L}$ and $\mathcal{P}(\tau)$ share the same eigenfunctions $l_k(\mathbf{x})$,
\begin{eqnarray}
    \mathcal{P}(\tau) l_k(\mathbf{x}) &=& \lambda_k(\tau)l_k(\mathbf{x})\, ,  
\label{eq:MSM_propagator_eigenvalues}    
\end{eqnarray}
and the propagator eigenvalues $\lambda_k(\tau)$ are related to $\kappa_k$, the eigenvalues of $\mathcal{L}$, as $\lambda_k(\tau) = \text{e}^{\kappa_k \tau}$.
The leading eigenvalues and eigenfunctions provide important insights into the microscopic dynamics, such as identifying long-lived conformational states and the transition rates between them.
\begin{eqnarray}
   t_k^{\mathrm{ITS}}  &=& -\frac{\tau}{\ln\lambda_k(\tau)} \quad \forall\tau > 0\, .
\label{eq:MSM_its}   
\end{eqnarray}

To compute these eigenfunctions and eigenvalues, the propagator is discretized with respect to a set of $n$ ansatz functions: $\chi_i: \mathbb{R}^m \rightarrow \mathbb{R}$.
Eq.~\ref{eq:MSM_propagator_eigenvalues} is then represented as a generalized matrix eigenvalue equation
\begin{eqnarray}
    \mathbf{C}(\tau) \mathbf{l}_k &=& \lambda_k(\tau) \mathbf{S}\,\mathbf{l}_k
\end{eqnarray}
where the matrix elements are (time) correlation functions of the ansatz functions, with the MSM lag time being $\tau$:
\begin{eqnarray}\label{eq:MSM_corr}
    C_{ij}(\tau) &=&  \lim_{T \to \infty}\int_0^T \chi_i(t)\chi_j(t+\tau) \,\mathrm{d}t \cr
    S_{ij} &=&  \lim_{T \to \infty}\int_0^T \chi_i(t)\chi_j(t) \,\,\,\mathrm{d}t    
\end{eqnarray}
The elements of $\mathbf{l}_k$ are related to the eigenfunctions as
$[\mathbf{l}_k]_i = \int l_k(\mathbf{x}) \mu(\mathbf{x}) \chi_i(\mathbf{x})\, \mathrm{d}\mathbf{x}$, where $\mu(\mathbf{x})$ is the stationary density of the overdamped Langevin dynamics,~i.e. the Boltzmann distribution.
In variational MSMs \cite{nuske2014variational}, the ansatz functions can be arbitrary functions of the CV space. 
In conventional MSMs \cite{prinz2011markov}, the discretization of the propagator is constructed by discretizing the CV space into $n$ non-overlapping  microstates $\Omega_i$, such that the entire CV space is covered: $\bigcup_{i=1}^n \Omega_i = \mathbb{R}^m$. 
The ansatz functions are then the microstate indicator functions
\begin{eqnarray}
    \chi_i(\mathbf{x}) &=&  \begin{cases} 
                            1   &\mbox{if } \mathbf{x} \in S_i \\
                            0   &\mbox{otherwise.}
                            \end{cases}    
\end{eqnarray}

\section{Methods}
\subsection{Simulations}

We benchmarked thermostat effects across a broad set of molecular systems using classical MD (GROMACS\cite{GROMACS}) and ab initio MD (CP2K\cite{A317,CP2K}). For water (TIP3P, TIP4P) and three organic liquids (pentane, aniline, glycerol), we generated equilibrated simulation boxes and performed extensive NVE and NVT production runs. Dynamical properties were evaluated under commonly used deterministic and stochastic thermostats—velocity-rescale, Berendsen, Nosé–Hoover, and stochastic dynamics—covering coupling times from 0.001 to 100 ps.
To assess a wide range of dynamical observables, we computed diffusion coefficients, shear viscosities, kinetic-energy distributions, and vibrational densities of states (vDOS). The latter were obtained from CP2K Born–Oppenheimer MD simulations using the semi-empirical GFN1-xTB method. For pentane, we also constructed Markov state models.
An overview of all simulation setups and parameters is provided in the supplementary information.

\subsection{Analysis}
\paragraph{Energy distribution.}
For the energy distributions we divide the total energy of the system of every frame by the number of molecules in the system. The values are then sorted via a basic histogram function. For the NVT plot in Fig.~\ref{fig:energyDistribution} a basic normal distribution was used. $\sigma^2$ is calculated with Eq.~\ref{eq:sigma}.
\paragraph{Diffusion constant.}
The Wiener-Kinchin \cite{Wiener-Kinchin} theorem was used with Eq.~\ref{eq:GK-self-diffusion} to directly calculate the velocity auto-correlation function VACF from the center of mass averaged velocities. The results shown are always for a block average. The trajectory is cut into 5 200~ps long pieces. The correlation functions are an average of the three directions x,y and z for the VACF. A detailed implementation can be seen in the supplementary information of Heinz \textsl{et al.}\cite{Heinz24}. For the final results, integration of the ACF stops at the plateau which is at 10~ps.
\paragraph{Shear viscosity.}
The Wiener-Kinchin \cite{Wiener-Kinchin} theorem was used with Eq.~\ref{eqn:GK_viscosity} to calculate the pressure tensor auto correlation function PACF from the pressure tensor. 
The results shown are always for a block average. The trajectory is cut into 20 1~ns long pieces. The correlation function are the average of the xy,xz and yz pressure components for the PACF. For the final results integration of the ACF stops at the plateau. This is 10~ps for water viscosity. 100~ps for anilin and pentane and 5~ns for glycerol.
\paragraph{Vibrational Density of States.}
The vibrational density of states (vDOS) was obtained as the Fourier transform of the discrete time-correlation function (TCF) of the atomic velocities taken from the AIMD simulations (\textit{cf.} Eq.~\ref{eq:vdos}), using the \textit{ChirPy} python package (version 0.27.1), available on GitHub.\cite{ChirPy,GitHub_ChirPy}
A triangular filter was applied to the TCF in combination with a Welch window function with a length of 25~ps in order to account for finite-size effects. 
For better visualization, the spectra were smoothed with a Gaussian filter setting $\sigma$=5~cm$^{-1}$.
\paragraph{Markov state models.}
Markov state models are constructed for the liquid dynamics of pentane. MD trajectories are realized with the GROMACS stochastic dynamics integrator ($\tau_T^{GSD} = 0.001\, \mathrm{ps}, 0.01\, \mathrm{ps}, 0.1\, \mathrm{ps}, 1\, \mathrm{ps}, 10\, \mathrm{ps}, 100\, \mathrm{ps}$) and the leap frog integrator with different thermostats (VR ($\tau_T^{VR} =1\, \mathrm{ps}$), B ($\tau_T^{B} = 0.1\, \mathrm{ps}$) and Nosé-Hoover (NH) thermostat ($\tau_T^{NH} = 4\, \mathrm{ps}$). The detailed MD simulation methodology can be found in Sec.~\ref{SI:SI:Theory}-E.
The torsion angles ($\phi$ (H-C2-C3-H) and $\psi$ (H-C3-C4-H)) of $140$ pentane molecules are written out every $100$~$\mathrm{fs}$ as MD output. Five data sets of equal size, each containing $28$ molecules and spanning $10$~ns, are used to analyze the dynamic observables within the standard deviation between these sets.
From each torsion angle trajectory, we constructed two-dimensional MSMs on a $3 \times 3$ grid, with the interval $[-180, -70, 70, 180]$ along both dimensions. The nine microstates correspond to the maximum values of the ($\phi$,$\psi$)-angle distribution, cf. Fig.~\ref{SI:SI_fig:MSM}. 
The lag time was varied between $0.1$~ps and $59.1$~ps in steps of $1$~ps.

\section{Results}
We calculate time-correlation functions that decay on time scales ranging from $10^{-2}$ ps (velocity autocorrelation functions in water) to $10^3$ ps (pressure autocorrelation function in viscous liquids).
We examine the influence of deterministic and stochastic thermostats on the decay time of correlation functions and on the derived observables, such as the diffusion constant, viscosity, vDOS and MSM implied timescales.
The thermostat coupling times are varied between 0.001 ps and 100 ps, with each coupling time represented by the same color throughout the manuscript.

%
%
\subsection{Diffusion coefficient of water}
\begin{figure}[h]
    \centering
    \includegraphics[width=\columnwidth]{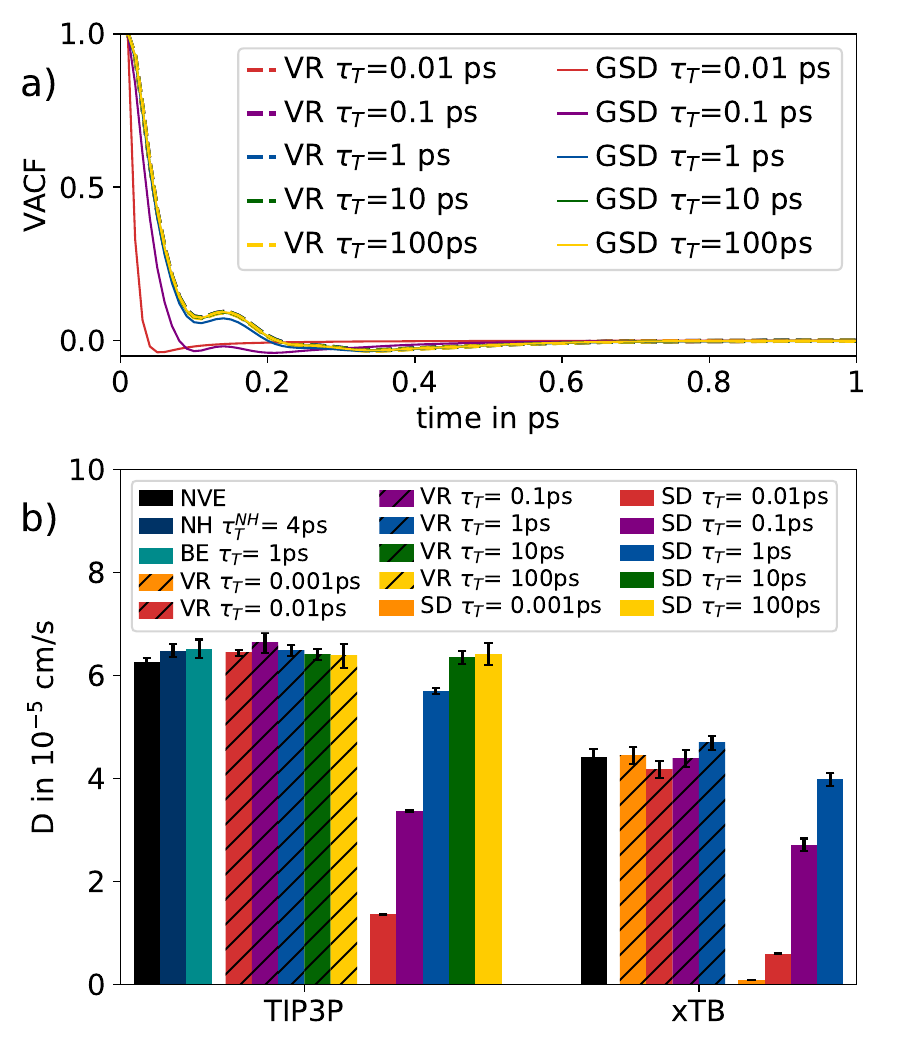}    
    \caption{Self diffusion  of water. 
    a) VACF of TIP3P water for different thermostat coupling times $\tau_T$ of VR thermostat and SD thermostat.
    b) Apparent self-diffusion coefficient of TIP3P and xTB water for various thermostats and thermostat coupling times $\tau_T$. Error bars show standard deviations obtained from block averaging.
    }
    \label{fig:TIP3P_VACF}
\end{figure}
Based on both on  empirical and ab-initio MD simulations, we calculated the diffusion coefficient of water from the molecular VACF (Eq.~\ref{eq:GK-self-diffusion}).
The VR thermostat did not alter the VACF compared to deterministic thermostats and the NVE reference simulation, even when the VR coupling time was varied (Fig.~\ref{fig:TIP3P_VACF}.a and \ref{SI:fig:TIP3P_VACF}).
By contrast, the GSD thermostat distorted the VACF for short coupling times ($\tau_T < 1\, \mathrm{ps}$), introducing a faster decay relative to NVE.
For moderate and large coupling times ($\tau_T \ge 1\, \mathrm{ps}$), however, the VACF progressively approached the NVE reference. 
Due to the fast decay of the VACF at short coupling times, the apparent diffusion coefficient is lower than the NVE reference (Fig.~\ref{fig:TIP3P_VACF}.b).
The same results are observed for the TIP4P water model (Fig.~\ref{SI:fig:TIP4P_VACF}).
The effect of the GSD thermostat is consistent with observations in the literature: a strongly coupled stochastic thermostat (i.e., low $\tau_T$) considerably slows down the self-diffusion of water\cite{Junghans2007, basconi2013effects, Ruiz-Franco2018}.
The same trend can be reproduced using the xTB water model and the OBABO thermostat (Fig.~\ref{SI:fig:xTB}).
Setting $\tau_T=0.001$~ps leads to a  significantly reduced apparent diffusion constant of $D=0.09~10^{-5}$~cm$^2$~s$^{-1}$, compared to a $D=3.98 10^{-5}$~cm$^2$~s$^{-1}$ at $\tau_T=1$~ps.

We report apparent diffusion coefficients without applying finite-size corrections for periodic boundary condition\cite{Yeh2004}. 
The much smaller diffusion constant of xTB water ($D=4.41 \cdot 10^{-5}$~cm$^2$~s$^{-1}$ for NVE) compared to TIP3P water  ($D=6.25 \cdot10^{-5}$~cm$^2$~s$^{-1}$ for NVE)  is partly due to xTB being a more accurate water model (experimental value for water at 25 $^{\circ}$C: $D=2.299 \cdot 10^{-5}$~cm$^2$~s$^{-1}$), and partly due to finite size effects of the smaller xTB simulation box.
(See also Tab.~\ref{SI:SItab:results_water}).

%
%
\subsection{Shear viscosity} 
\begin{figure}[h]
    \centering
    \includegraphics[width=\columnwidth]{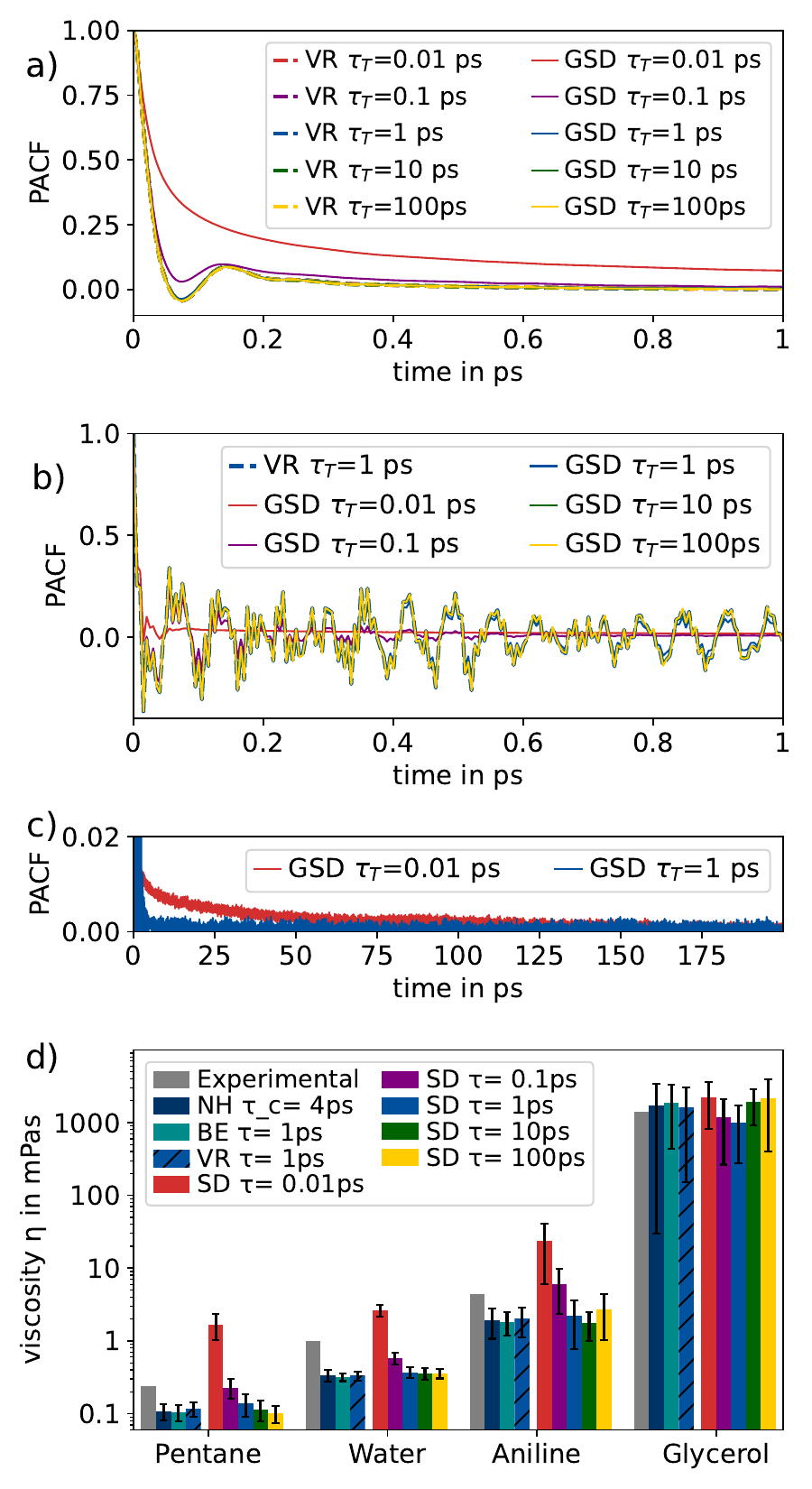}
    \caption{Shear viscosity. PACF of TIP3P water (a) and anilin (b) for different thermostat coupling times $\tau_T$ of VR thermostat and SD thermostat. 
    c)Long tail of the PACF of anilin.
    d) Shear viscosities of pentane, TIP3P water, anilin and glycerol for different thermostats and coupling times $\tau_T$. The error is the standard deviation obtained from block averaging. Note the logarithmic scale of the viscosity-axis. 
   }
    \label{fig:viscosity_result}
\end{figure}

The previous section showed that stochastic thermostats with coupling times below 1 ps significantly distort the VACF. 
Next, we explore whether the influence of the stochastic thermostat is detectable for autocorrelation functions whose decay times far exceed typical thermostat coupling times. 
To this end, we calculate shear viscosities from the atomic PACF (Eq.~\ref{eqn:GK_viscosity}) for liquids with markedly different viscosities. 
Specifically, we examine pentane, water, anilin and glycerol whose viscosities span four order of magnitude, ranging from 0.1 mPas to 1400 mPas.\cite{viscosities}
Accordingly, the PACFs of these substances exhibit widely varying decay times, from ~1 ps for water to hundreds of picoseconds for aniline (Fig.~\ref{fig:viscosity_result}.c) and several nanoseconds for glycerol.
The pressure in Eq.~\ref{eqn:GK_viscosity} can be calculated using either atomic velocities and forces (atomic approach) or from the center-of-mass quantities (molecular approach). 
While atomic PACFs exhibits strong oscillations due to intramolecular vibrations, molecular PACFs decays more smoothly, typically showing only a single oscillation.
Despite these differences, both PACFs yield equivalent results when integrated, and thus either approach can be used to calculate the viscosity of a liquid. \cite{Cui96}.
In our analysis, we use the atomic PACFs, which show the typical fluctuations (Fig.~\ref{fig:viscosity_result}.b, \ref{SI:SIfig:Pentane}, \ref{SI:SIfig:Anilin}, \ref{SI:SIfig:Glycerine}).
However, for a relatively rigid molecule, the atomic PACF reduces the to molecular PACF leading to a smooth appearance (Fig.~\ref{fig:viscosity_result}.a, \ref{SI:fig:TIP3P_PACF}, \ref{SI:fig:TIP4P_PACF}).
Similar to VACFs, PACFs obtained using the VR thermostat are indistinguishable from those of the NVE reference simulation, regardless of the thermostat coupling time. This confirms the VR thermostat does not affect the PACF and does not bias the viscosity estimate \cite{basconi2013effects}.
In contrast, the stochastic GSD thermostat distorts the PACF when the coupling time $\tau$ is shorter than 1 ps (Figs.~\ref{fig:viscosity_result}, 
\ref{SI:fig:TIP3P_PACF}, \ref{SI:fig:TIP4P_PACF}, \ref{SI:SIfig:Pentane}, \ref{SI:SIfig:Anilin}, \ref{SI:SIfig:Glycerine}). 
The distortion is most pronounced at $\tau = 0.01$ ps. At $\tau = 0.1$ ps, the PACF begins to resemble the NVE reference, and by $\tau = 1$ ps, the GSD PACFs agree well with the reference.
As a result, viscosities are substantially overestimated in GSD simulations with $\tau = 0.01$ ps and moderately overestimated at $\tau = 0.1$ ps. For $\tau \geq 1$ ps , the computed viscosities match the NVE values well (Fig.~\ref{fig:viscosity_result}~d), Tab.~\ref{SI:SItab:results_water}). The only exception is glycerol, where the viscosities obtained across different coupling times agree within statistical uncertainty due to its high viscosity.
These findings demonstrate that, for stochastic thermostats, the influence of the thermostat coupling time persists even for slowly decaying autocorrelation functions, and is not limited to fast dynamical processes.
Next, we examine whether the influence of the thermostat coupling time extends to properties that are not straightforward integrals of correlation functions, but are instead derived from more complex mathematical expressions.
Specifically, we consider the vibrational density of states (vDOS), obtained via a Fourier transform  (Sec.~\ref{sec:vDOS_results}) and MSM slow processes, computed from the eigenvectors and eigenvalues of a (row-normalized) correlation matrix (Sec.~\ref{sec:MSM_results}).
%

%
%
\subsection{Vibrational Density of States of Water}
\label{sec:vDOS_results}

\begin{figure}[h]
    \centering
    \includegraphics[scale=0.56]{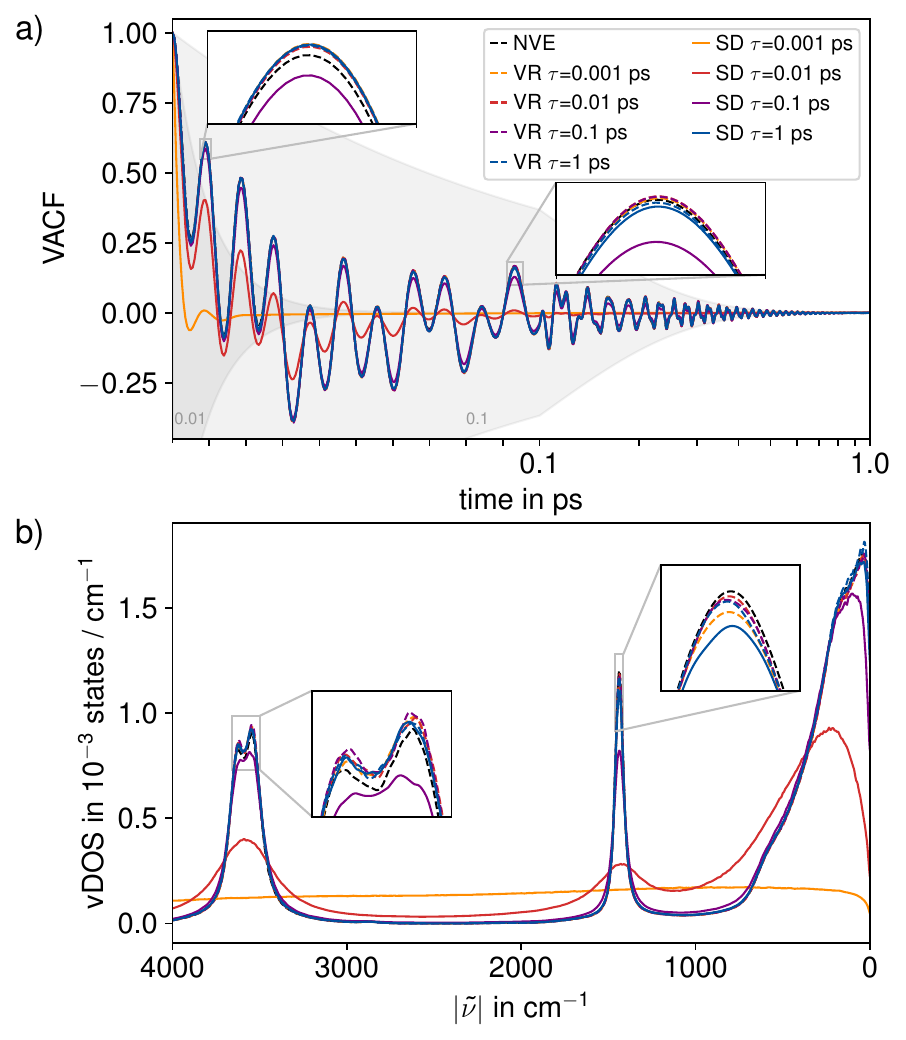}
    \caption{a) Atomic vibrational autocorrelation function (VACF) of liquid water obtained from AIMD simulations using various coupling times $\tau_T$ for both velocity rescaling (VR) and stochastic dynamics (SD, here: OBABO). A logarithmic scale is used for lag times greater than 0.1~ps. The shaded areas represent the envelopes of ideal damped oscillators with decay times  of $\tau_T=0.01$ and $0.1$~ps, respectively.
    b) Vibrational density of states (vDOS), computed as the Fourier transform of the VACF.}
    \label{fig:xTB_vDOS}
\end{figure}

Based on AIMD simulations, we calculated the vibrational density of states (vDOS) of liquid water using Eq.~\ref{eq:vdos}.
In all simulation setups, the time correlation function decays rapidly within $1$~ps, with most of the vibrational signatures captured within the first $0.1$~ps.
We found that the effect of the VR thermostat on the atomic VACF was negligible compared to the NVE reference, as shown in Fig.~\ref{fig:xTB_vDOS}.a). 
Even for very short coupling times, VR does not affect the correlation of the individual degrees of freedom. This observation is consistent with the findings reported by Bussi \textit{\textsl{et al.}} in their original publication on the VR thermostat.\cite{bussi2008stochastic}
In contrast, we see clear differences in Fig~\ref{fig:xTB_vDOS}.a for SD, where the stochastic term interferes with every degree of freedom, leading to damped oscillations in the atomic velocities and a rapid randomization of the dynamics, up to a complete destruction of the VACF for $\tau_T=0.001~$ps. 
However, this influence diminishes with a coupling constant of $0.1~$ps and disappears with $1$~ps, allowing the system to closely reproduce the NVE reference. 
The VACF is damped by two effects: intrinsic molecular friction and the influence of the thermostat. The effect of the thermostat can be likened to that of an ideal damped oscillator, where the decay constant corresponds to the thermostat coupling time.
Comparing the VACF to an ideal damped oscillator, as shown in Fig.~\ref{fig:xTB_vDOS}.a, we find that intrinsic damping occurs faster than $\tau_T = 0.1$~ps.
Consequently, the overall damping from the stochastic thermostat
OBABO with coupling strengths in this range is only slightly greater.
For $\tau_T = 0.01$~ps the oscillations found exceed the envelope function corresponding to an ideal damping of this strength, driven by the underlying atomic correlations.
This behavior shows the robustness of the system’s intrinsic dynamics, which partially counteract the imposed stochastic damping.

The Fourier transformation of the atomic VACF yields the vibrational density of states (vDOS) as defined in Eq.~\ref{eq:vdos}, and shown for liquid water in Fig.~\ref{fig:xTB_vDOS}.b). All simulation setups except for SD with $\tau_T=0.001$~ps reproduce the vibrational bands in the regions
0--1000~cm$^{-1}$ (libration), 1400--1600~cm$^{-1}$ (bending) and 3500--3700~cm$^{-1}$ (stretching).
The use of the xTB method results in slightly shifted peak positions compared to the experiment.\cite{Bertie1996a,Silvestrelli1997}
In line with what we observed for the VACF, VR does not significantly change the spectrum. However, there are clear effects for SD with short coupling times.
The peak widths increase with decreasing $\tau_T$, which is the expected behavior of a damped oscillator (the Fourier transform of the damping function $\mathrm{e}^{-t/\tau_T}$ is a Lorentzian whose width is proportional to $1/\tau_T)$.
While the vibrational bands remain identifiable at $\tau_T=0.01$~ps, all vibrational information is lost at $\tau_T=0.001$~ps.
It should be noted, however, that the equipartition theorem remains satisfied, and the vDOS integrates to 1 for any chosen value of $\tau_T$.

\subsection{Markov state model}
\label{sec:MSM_results}
\begin{figure*}[t]
    \centering
 \includegraphics[width=\textwidth]{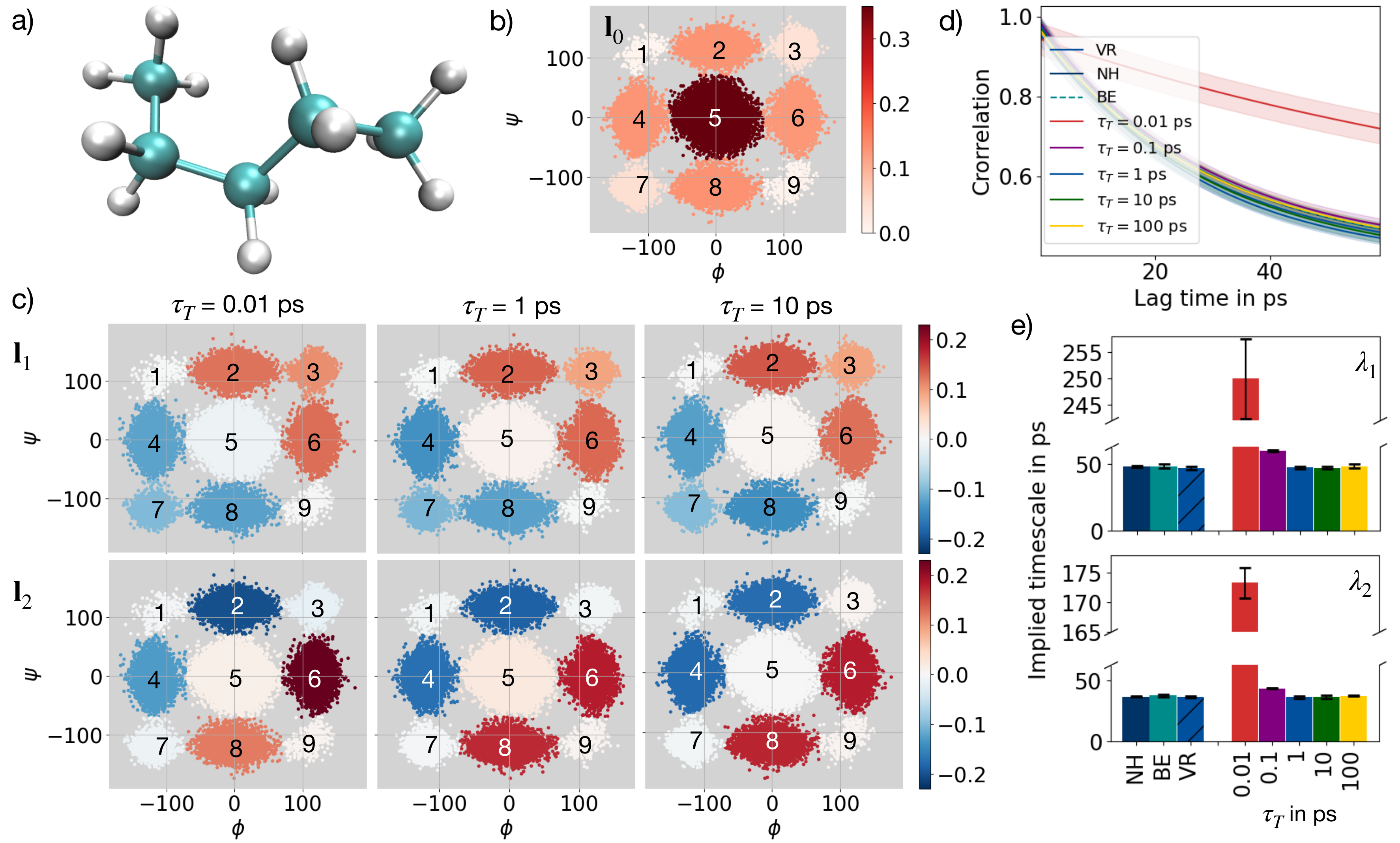}
    \caption{
    Markov state model for a) pentane.
    b) Stationary distribution at lag time $\tau=20.1$~ps. 
    c) Slowest dynamical processes ($\mathbf{l}_1$, $\mathbf{l}_2$) for GSD with different coupling times $\tau_T=0.01$~ps,  $\tau_T=1$~ps and $\tau_T=10$~ps, with shared color bar.
    d) MSM correlation matrix elements $C_{5,5}$ as function of the lag time ($0.1-59.1$~ps).
    e) MSM implied timescales as mean with standard deviation over converged lag time ($20.1-59.1$~ps) for $\lambda_1$ and $\lambda_2$.
    }
    \label{Fig-MSM}
\end{figure*}

We compare the effect of the different thermostats on the results of the MSM time-lagged correlation function for pentane, Eq.~\ref{eq:MSM_corr}.
The first MSM eigenvector represents the stationary distribution, as shown in Fig.~\ref{Fig-MSM}.b.
The largest metastable state (cluster 5) corresponds to the trans-trans conformation of pentane.
The next most populated clusters are 2, 4, 6 and 8, corresponding to the four possible trans-gauche and gauch-trans conformations. Even more weakly populated are gauche-gauche conformations in clusters 1, 3, 7 and 9. 
Each MD set up, yields the same first MSM eigenvectors, indicating 
that the stationary properties are independent of the choice of thermostat and thermostat coupling time.

The dynamic information in an MSM consists of two parts. 
First, the eigenvectors of the MSM transition matrix encode transitions or a set of transitions, where the dominant eigenvectors represent the slow transitions in the system.
Second, the eigenvalues encode the implied timescale at which the transitions occur. 
The pentane MSM exhibits excellent convergence of the implied timescales, enabling accurate determination of dynamic timescales.
The MSM exhibits a slow process at $t_1^{\mathrm{ITS}} = 47\, \mathrm{ps}$, followed by a three processe with similar implied timescales 
$t_2^{\mathrm{ITS}} = 37\, \mathrm{ps}$, 
$t_3^{\mathrm{ITS}} = 34\, \mathrm{ps}$, 
$t_4^{\mathrm{ITS}} = 32\, \mathrm{ps}$ and
$t_5^{\mathrm{ITS}} = 21\, \mathrm{ps}$ 
(Fig.~\ref{SI:SI_fig:MSM}).
We find that not only the stationary eigenvector but also the dominant eigenvectors are independent of the choice of thermostat and thermostat coupling time.
Most notably, in simulations with the stochastic thermostat GSD, the dominant eigenvectors are unaffected by strong coupling constants $\tau_T \le 1\, \mathrm{ps}$ (Fig.~\ref{Fig-MSM}.c).
This indicates that the type of transitions, i.e. the conformational mechanisms, are not affected by the thermostat coupling times. 
By contrast, the implied timescales associated to these processes are distorted by strong GSD thermostat coupling times (Fig.~\ref{Fig-MSM}.e). 
Specifically, for GSD with $\tau_T=0.01$~ps the slow implied timescale of the slowest process is overestimated by about a factor 5 ($t_1^{\mathrm{ITS}} = 244.6$~ps). 
With $\tau_T=0.1$~ps, the implied timescale is moderately overestimated by 25\% ($t_1^{\mathrm{ITS}} =59.6$~ps), while at $\tau_T=1$~ps the implied timescale matches the NVE result.
This result can be traced back to the decay of the matrix elements in the MSM correlation matrix. 
Fig.~\ref{Fig-MSM}.f shows the evolution of the matrix element $C_{5,5}$ of the $9\times~9$ correlation matrix. 
This correlation functions represents self-transitions from cluster 5 to cluster 5 within lag time $\tau$, where the system may either leave the cluster and return, or stay within cluster 5 throughout the lag time.
Once again, the results of the particularly strongly coupled GSD data ($\tau_T=0.01$~ps) stand out. 
We observe a significantly weaker drop in the correlation function as a function of lag times than for all other thermostat coupling times.

\subsection{Error of the ACF for SD thermostats}
\begin{figure}[h]
    \includegraphics[width=\columnwidth]{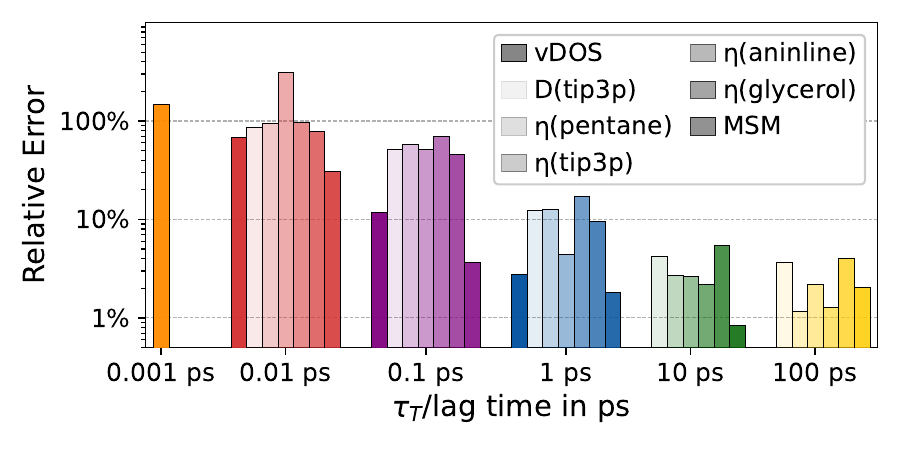}
    \caption{Relative error of the of integral over the auto-correlation functions with an SD thermostat compared to NVE for vDOS and D and compared to VR with $\tau_T=1$~ps for $\eta$ and MSM. The values for vDOS go from 0.001 ~ps to 1~ps and the other from 0.01~ps to 100~ps for the coupling time $\tau_T$.}
    \label{fig:SD_error}
\end{figure}

Throughout the study we found that stochastic thermostats with strong coupling constants of $\tau_T < 1\, \mathrm{ps}$ distort dynamic properties. 
However, for moderate coupling strengths of $\tau_T = 1\, \mathrm{ps}$, the resulting dynamics are in good agreement with those from NVE simulations. 
To quantify this deviations, we calculated the relative error of the various ACF with respect to the NVE simulation
\begin{equation}
    \delta= \frac{\int |\mathrm{ACF}(\tau) - \mathrm{ACF}(NVE)|\,\mathrm{d}\tau}{\int |\mathrm{ACF}(NVE)|\,\mathrm{d}\tau} \, , 
\end{equation}
where the integrals is only evaluated over the first picosecond for the diffusion and viscosity calculations. For the vDOS we use the complete 25~ps and for MSM all lag times.
Fig.~\ref{fig:SD_error} summarizes the results.
For very short coupling times, such as $\tau_T = 0.01\, \mathrm{ps}$, the relative error reaches approximately 100\%. 
While large, this is still within the same order of magnitude as the NVE reference simulations. 
As the coupling time increases, the error decreases significantly: at $\tau_T = 1\, \mathrm{ps}$, the relative error drops to 10\% or less. 
For coupling times of $\tau_T \geq 10\, \mathrm{ps}$, the error falls to around 1\%, which is below the typical statistical uncertainty of a dynamic observable.
(In the case of Markov state models (MSMs), it is worth noting that a relative error in the ACF integral does not result in a proportional error in the implied timescales.)

However, when the thermostat is coupled too weakly, the simulation may no longer sample the canonical ensemble accurately. 
For instance, in water simulations with $\tau_T = 100\, \mathrm{ps}$, the energy distribution deviates noticeably from the expected canonical distribution (see Fig.~\ref{fig:energyDistribution}). 
As a result, stationary properties may not be reliably reproduced.
%

\section{Summary and conclusion}
We evaluated the impact of commonly used thermostats - deterministic thermostat, velocity rescale thermostat, and stochastic thermostats - on time-correlation functions and the dynamic properties derived from them in molecular dynamics simulations.
Microcanonical (NVE) simulations serve as a reference. 
Consistent with prior studies, we found that strongly coupled stochastic thermostats significantly distort dynamic observables, while deterministic thermostats and the velocity rescale thermostat closely match the NVE results. 
Notably, this behavior holds across all four properties tested - diffusion, viscosity, vibrational spectra, and Markov state models - and across the full range of correlation function decay times of these properties. 
This demonstrates that stochastic thermostat effects are not confined to fast dynamics: they systematically impact dynamic properties, even when the thermostat coupling time is well separated from the system's intrinsic timescales.
To clarify this, consider that in NVE simulations, stochasticity arises only from interactions with surrounding particles. 
Stochastic integrators, however, introduce an additional thermostat-induced noise term. 
In solvated systems, environmental coupling occurs on $\sim$1~ps timescales.
Our results indicate that for $\tau_T = 1~\mathrm{ps}$ each time-correlation function is effectively governed by solvent-induced fluctuations, closely matching the NVE reference.
However, whether a strong stochastic thermostat accelerates or slows down the decay of a time-correlation function, depends on the particular function.
It accelerates the decay of VACFs, which affects quantities like the diffusion constant and the vibrational density of states. 
This is readily explained by the fact that stochastic thermostats apply random forces directly to particle velocities, thereby rapidly randomizing them.
By contrast, a strong stochastic thermostat slows down correlation decay in the PACF and in state-to-state correlation functions used in Markov state models (MSMs). 
This leads to overestimation of viscosity from the PACF, and to slower apparent transition rates and larger implied timescales in MSMs.
In MSMs, a position-dependent correlation function is evaluated. 
A stronger coupling in stochastic thermostats implies both increased random forces and enhanced friction, as required by the fluctuation-dissipation theorem. 
This high friction suppresses inertial motion, preventing the molecule from accumulating enough momentum to cross energy barriers efficiently. 
Hence, barrier crossings rely solely on random kicks, leading to lower state-to-state transition rates and longer implied timescales.
However, in the PACF, the pressure tensor elements are proportional to the kinetic energy as well as to the virial. 
Thus, this correlation function depends on the velocities as well as on the positions and forces through the virial. 
Empirically, we find that a strong stochastic thermostat slows down the decay of the PACF, leading to the observed overestimate of the corresponding viscositites. 
%

%
%
%
%

%
While stochastic thermostats do affect dynamics at short coupling times, the impact remains moderate. 
At a very short coupling time of $\tau_T = 0.01~\mathrm{ps}$,  relative errors in dynamic properties reach approximately 100\% compared to NVE reference data.
However, at coupling times of $\tau_T = 1~\mathrm{ps}$, which are commonly used in MD simulations, these errors drop to 10\% and fall further to 1\% for longer coupling times.
Although this may suggest that weak coupling is always preferable, very large $\tau_T$-values can lead to inadequate temperature control and the thermostat fails to maintain a proper NVT ensemble.
Thus, coupling times in the range of 1–10 ps for stochastic thermostats balance accuracy in dynamic properties with proper ensemble sampling.

\section{Acknowledgements}
Funded by the Deutsche Forschungsgemeinschaft (DFG, German Research Foundation):
CRC 1449 – project-ID 431232613, sub-project C02;
CRC 1114 - project-ID 235221301, sub-project B05;
GRK 2473 - project-ID 392923329, sub-project C1.
The authors would like to thank the HPC Service of FUB-IT, Freie Universität Berlin, for computing time.

\section{Supporting Information}
The supporting information contains a table for all simulated systems, further theory about correlation functions and MSM, as well as plots for ACF, its integral and numerical results for TIP3P, TIP4P, xTB, pentane, anilin, and glycerol. The analysis code can be found under https://github.com/bkellerlab/CorrelationFunctions\\\_and\_Thermostats.git .

\section{References}
\bibliography{bib}

\appendix

\newpage
\begin{figure}
    \centering
    \includegraphics[width=8cm]{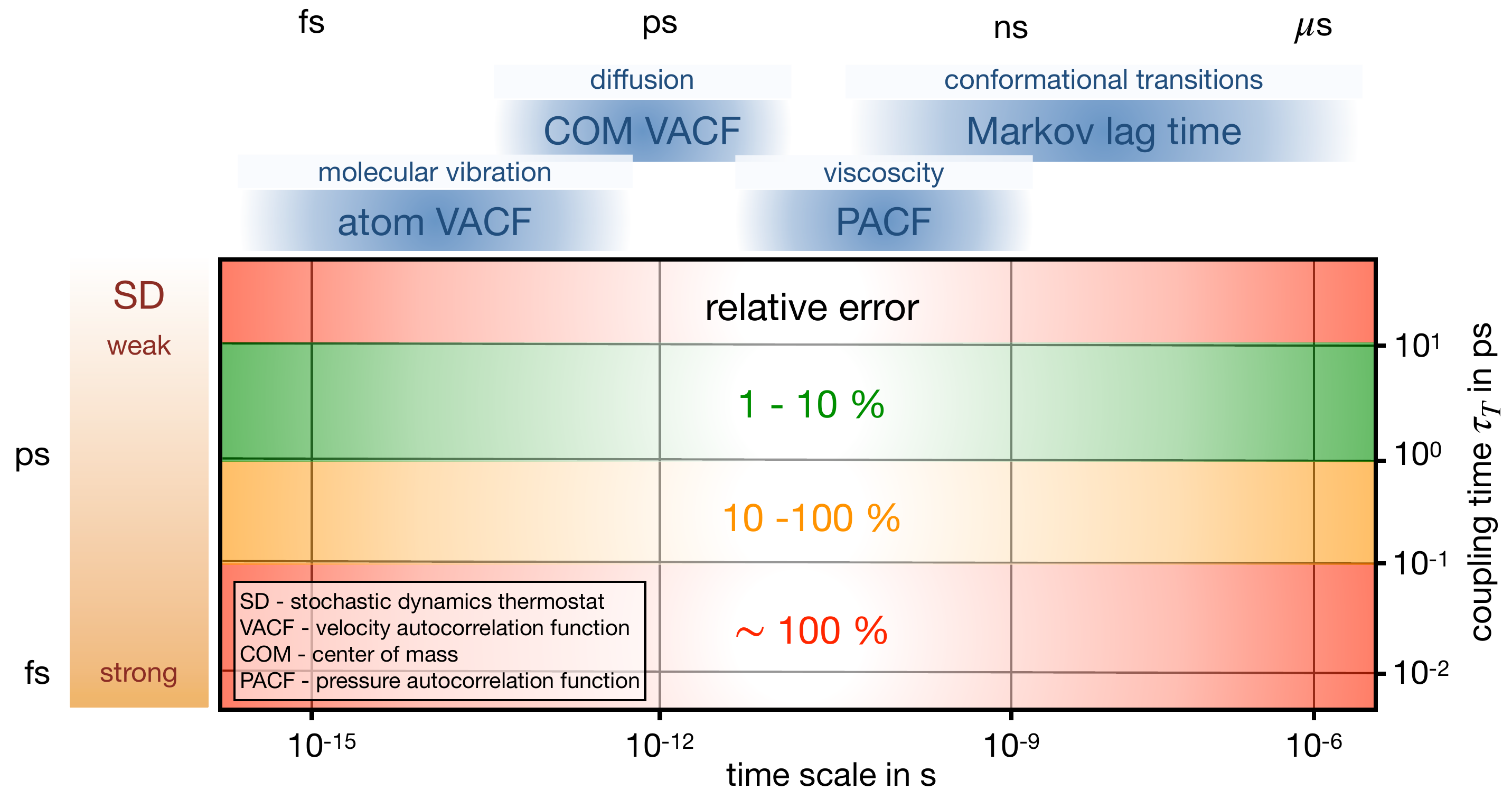}
    \caption{\textbf{Table-of-content graphic}}
    \label{fig:TOC}
\end{figure}

\newpage
\onecolumngrid

\renewcommand{\thesection}{SI: \Roman{section}}
\renewcommand{\thesubsection}{\thesection-\Alph{subsection}}
\renewcommand{\thefigure}{SI-\arabic{figure}}
\renewcommand{\thetable}{SI-\roman{table}}
\renewcommand{\theequation}{SI-\arabic{equation}}


\title{How Thermostats Influence Dynamics Across Time Scales: A Systematic Study from Fast Motions to Slow Transitions}
\author{Frederick Heinz}
\affiliation{ 
Freie Universität Berlin, Department of Biology, Chemistry and Pharmacy, Arnimallee 22, 14195 Berlin}

\author{Sascha Jähnigen}
\affiliation{ 
Freie Universität Berlin, Department of Biology, Chemistry and Pharmacy, Arnimallee 22, 14195 Berlin}

\author{Joana-Lysiane Schäfer}
\affiliation{ 
Freie Universität Berlin, Department of Biology, Chemistry and Pharmacy, Arnimallee 22, 14195 Berlin}

\author{Bettina G.~Keller}%
\email{bettina.keller@fu-berlin.de}
\affiliation{ 
Freie Universität Berlin, Department of Biology, Chemistry and Pharmacy, Arnimallee 22, 14195 Berlin}

\maketitle
\newpage

\section{Simulations}
\label{SI:Simulations}

\subsection{Water simulations with GROMACS}
\label{SI:method:MD}
We used GROMACS 2023\cite{GROMACS} to generate the starting configurations of the TIP3P\cite{TIP3P} and TIP4P\cite{TIP4P} water simulations. 
We filled a 3x3x3~nm³ cubic box with 884 TIP3P\cite{TIP3P} water molecules. 
After a steepest-descent minimization, the box is equilibrated for 2 ns in the NPT ensemble with the velocity-rescale thermostat\cite{bussi2007canonical,bussi2008stochastic} at 300~K and $\tau_T=1~\mathrm{ps}$ and Parinello-Rahman barostat\cite{Parinello-Rahman} 1~bar and$\tau_P=2~\mathrm{ps}$, a reference pressure of $P_{\mathrm{ref}}=1 \mathrm{bar}$ and a compressibility of $\chi=4.5\cdot 10^{-5} \mathrm{bar}^{-1}$.
The final configuration had a box size of 2.97 x 2.97 x 2.97~nm³, corresponding to a density $1.08$ g cm$^{-3}$.
Subsequently, we equilibrated the box in the NVT ensemble with the velocity-rescale thermostat\cite{bussi2007canonical,bussi2008stochastic} at 300~K and $\tau_T=0.1~ps$ for 2 ns.
The same procedure was applied to generate the starting configuration for the TIP4P\cite{TIP4P} water simulations.
This yielded a box of 3.02 x 3.02 x 3.02~nm³, corresponding to a density $0.986$ g cm$^{-3}$.
For the equilibration runs, we used a leap frog integrator and periodic boundary conditions in all three spatial dimensions. 
The O-H bonds were constrained using the LINCS algorithm\cite{LINCS} (LINCS iteration = 1, LINCS order = 4), allowing for a timestep for 2 fs. 
Van-der-Waals interactions were cut off at 1.2~nm, and electrostatic interactions were calculated with the Particle-Mesh-Ewald algorithm \cite{PME} (order = 4, Fourier spacing = 0.16).

For the TIP3P box, we performed a set of production simulations to calculate the energy distribution, where we simulated the system for 60~ns with a time step of 2~fs. 
Energies were written to file very 1 ps. 
We performed NVE simulations using GROMACS 2021.4\cite{GROMACS} with double precision and the leap-frog integrator.
Additionally, we performed NVT simulations using GROMACS 2023 in single precision.
We used GROMACS stochastic dynamics\cite{goga2012efficient} integrator with $\tau_T^{GSD} = 2\, \mathrm{ps}$;
leap frog and Velocity Rescale (VR) thermostat \cite{bussi2007canonical,bussi2008stochastic} with $\tau_T^{VR} = 1\, \mathrm{ps}$;
leap frog and Berendsen (BE) thermostat\cite{Berendsen1984}  with  $\tau_T^{B} = 1\, \mathrm{ps}$; and 
leap frog and Nosé-Hoover (NH) thermostat \cite{Nose-Hover} with  $\tau_T^{NH} = 4\, \mathrm{ps}$ and a chain length of 1.
The thermostats VR and B were coupled to the system every 100 time steps, whereas NH was coupled every time step. 
Periodic boundary conditions, van-der-Waals interactions and electrostatic interactions were treated as in the equilibration runs.

For the TIP3P and TIP4P boxes, we performed two separate sets of productions simulations to determine the self-diffusion coefficient and the shear viscosity.
In the set for the self-diffusion coefficient, the systems were simulated for 1 ns.
Positions and velocities were written to file every 10 fs. 
In the set for the shear viscosity, the systems were simulated for 20 ns. 
The pressure tensor was written to file every 5 fs.
In each set, we varied the ensemble and the thermostats. We performed NVE simulations using GROMACS 2021.4\cite{GROMACS} with double precision and the leap-frog integrator.
Additionally, we performed NVT simulations using GROMACS 2023 in single precision.
We used GROMACS stochastic dynamics integrator with 
$\tau_T^{GSD} = 0.001\, \mathrm{ps}$, $0.01\, \mathrm{ps}$, $0.1\, \mathrm{ps}$, $1\, \mathrm{ps}$, $10\, \mathrm{ps}$, $100\, \mathrm{ps}$;
leap frog and Velocity Rescale (VR) thermostat with 
$\tau_T^{VR} = 0.001\, \mathrm{ps}$, $0.01\, \mathrm{ps}$, $0.1\, \mathrm{ps}$, $1\, \mathrm{ps}$, $10\, \mathrm{ps}$, $100\, \mathrm{ps}$;
leap frog and Berendsen (BE) thermostat with 
$\tau_T^{BE} = 0.1\, \mathrm{ps}$; and 
leap frog and Nosé-Hoover (NH) thermostat with 
$\tau_T^{NH} = 4\, \mathrm{ps}$ and a chain length of 1.
The thermostats VR and B were coupled to the system every 100 time steps, whereas NH was coupled every time step. 
We used a time step of 1~fs and did not constrain the O-H bonds.
Periodic boundary conditions, van-der-Waals interactions and electrostatic interactions were treated as in the equilibration runs. 
See Tab. \ref{SI:SI-simulations_water} in the SI for an overview of the simulations.

\subsection{Water simulations with CP2K}
\label{SI:method:aimd}
The AIMD simulations were performed as Born-Oppenheimer MD with CP2K (version 2024.1)\cite{A317,CP2K} and the Quickstep module\cite{G016} using the semi-empirical GFN1-xTB method developed by Grimme \textsl{et al.}\cite{M079}.
For the simulations the velocity-Verlet integrator was used together with periodic boundary conditions in all three spatial dimensions.
The CSVR thermostat\cite{bussi2008stochastic} was used to emulate both velocity rescaling (VR) and stochastic dynamics (SD), according to Eq.~\ref{eq:velocityRescale} with a global ($N_f=3\times\text{no. of particles}$) and massive ($N_f=1$) region, respectively.

A sample of 128 water molecules was considered in a cubic super cell of 15.672~\AA{}, which corresponds to a density of 0.9965~g~cm$^{-3}$.
The system was equilibrated for 20~ps at 300~K using a massive CSVR thermostat\cite{bussi2008stochastic} with a coupling constant $\tau_T=0.01$~ps (for the first 2.5~ps) and 0.1~ps.

Using the equilibrated configuration as a starting point, one microcanonical (NVE) and eight canonical (NVT) trajectories of 25~ps length were generated using a time step of 0.5~fs, with the trajectory written to file at each time step.
For the canonical runs varying thermostat setups were chosen using coupling constants $\tau_T=0.001, 0.01, 0.1, 1.0$~ps together with global (=VR) or massive (=SD using the OBABO splitting method) thermostatting at 300~K, respectively.

This procedure was repeated with four other statistically independent starting points that were sampled from one of the former NVT trajectories (global CSVR, $\tau_T=0.01$~ps), resulting in five trajectories for each MD setup (45 in total). All results shown are averages over the respective set of five trajectories.

See Tab. \ref{SI:SI-simulations_water}  for an overview of the simulations.

\subsection{Organic Liquids with GROMACS}\label{SI:MD_OrganicLiquids}
For the organic liquids anilin, pentane and glycerol a modified GROMOS\cite{GROMOS} force field was used. The parameters were obtained from the automated topology builder\cite{ATB}. The generation of the starting structures followed the same protocol as the for the TIP3P water boxes, where we filled an initial 3x3x3~nm³ cubic box with 191 aniline, 140 pentane or 600 glycerol molecules respectively. Furthermore, the equilibration runs lasted 10~ns to allow for better equilibration.
This resulted in a final simulation box for aniline of $3.13 \times 3.13 \times 3.13$~nm$^3$ and a density of $0.958$~g~cm$^3$. 
For pentane, the box is $2.93\times2.93\times2.93$~nm$^3$ resulting in a density of $0.669$~g ~cm$^{-3}$.
For glycerol, the box is $4.19\times4.19\times4.19$~nm$^3$ resulting in a density of $1.245$  g cm$^3$. 
For the production simulation only the simulations for the shear viscosity was performed (20 ns simulation time, pressure tensor written to file every 5 fs). We used 
GROMACS stochastic dynamics integrator with 
$\tau_T^{GSD} = 0.001\, \mathrm{ps}$, $0.01\, \mathrm{ps}$, $0.1\, \mathrm{ps}$, $1\, \mathrm{ps}$, $10\, \mathrm{ps}$, $100\, \mathrm{ps}$;
leap frog and Velocity Rescale (VR) thermostat with 
$\tau_T^{VR} =1\, \mathrm{ps}$;
leap frog and Berendsen (BE) thermostat with 
$\tau_T^{B} = 0.1\, \mathrm{ps}$; and 
leap frog and Nosé-Hoover (NH) thermostat with 
$\tau_T^{NH} = 4\, \mathrm{ps}$ and a chain length of 1.
The thermostats VR and B were coupled to the system every $100$ time steps, whereas NH was coupled every time step. 
We used a time step of $1$~fs and did not use bond constraints.
Periodic boundary conditions, van der Waals interactions, and electrostatic interactions were treated as in the equilibration runs for water.

See Tab. \ref{SI:SI-simulations_organic} in the SI for an overview of the simulations.

In addition to the above simulation, we have created a further pentane simulation from the equilibrated pentane box. For the MSM analysis, a simulation with 10~ns length and an output of the positions every 100~fs was performed.

\begin{table}[H]
\centering
\begin{small}
    \centering
    \begin{tabular}{l|l|| c|c || c|c ||c ||c}
    \hline              
            \multicolumn{7}{c||}{884 water molecules in $3\times3\times3$~nm$^3$ box } & 128 water molecules in (15.672~\AA{})$^3$ box  \\
    \hline  
            & &\multicolumn{2}{c||}{timestep: 1 fs}  
            & \multicolumn{2}{c||}{timestep: 1 fs}
            & timestep: 2 fs
            & timestep 0.5 fs\\
            & &\multicolumn{2}{c||}{output $x_i$/$v_i$: 10 fs }  
            & \multicolumn{2}{c||}{output $P_{xy}$: 5 fs}
            & output $E$: 1 ps
            & output $x_i$/$v_i$: 0.5 fs   \\
     MD&$\tau$ in ps & TIP3P & TIP4P & TIP3P & TIP4P & TIP3P & xTB\\
    \hline\hline
    LF & -  &1 ns&1 ns  &20 ns&20 ns    &60 ns\\ \hline\hline
    \multirow{6}{*}{GSD} 
    & 0.001 &1 ns&1 ns&20 ns&20 ns&\\ \cline{2-8}
    & 0.01  &1 ns&1 ns&20 ns&20 ns&\\ \cline{2-8}
    & 0.1   &1 ns&1 ns&20 ns&20 ns&\\ \cline{2-8}
    & 1     &1 ns&1 ns&20 ns&20 ns&\\ \cline{2-8}
    & 2     &&&&&60 ns\\ \cline{2-8}
    & 10    &1 ns&1 ns&20 ns&20 ns&\\ \cline{2-8}
    & 100   &1 ns&1 ns&20 ns&20 ns&60 ns\\ \hline\hline
    \multirow{4}{*}{SD$^*$} 
    & 0.001 &&&&&&25 ps\\ \cline{2-8}
    & 0.01  &&&&&&25 ps\\ \cline{2-8}
    & 0.1   &&&&&&25 ps\\ \cline{2-8}
    & 1     &&&&&&25 ps\\ \hline\hline
    \multirow{6}{*}{VR} 
    & 0.001 &1 ns&1 ns&20 ns&20 ns&&25 ps\\ \cline{2-8}
    & 0.01  &1 ns&1 ns&20 ns&20 ns&&25 ps\\ \cline{2-8}
    & 0.1   &1 ns&1 ns&20 ns&20 ns&&25 ps\\ \cline{2-8}
    & 1     &1 ns&1 ns&20 ns&20 ns&60 ns &25 ps\\ \cline{2-8}
    & 10    &1 ns&1 ns&20 ns&20 ns&\\ \cline{2-8}
    & 100   &1 ns&1 ns&20 ns&20 ns&\\ \hline\hline
     \multirow{2}{*}{BE} 
     & 0.1 &1 ns&1 ns&20 ns&20 ns&\\ \cline{2-8}
     & 1   &&& &&60 ns\\ \hline \hline
    NH & 4 &1 ns&1 ns&20 ns&20 ns&60 ns\\ \hline
    \end{tabular}
\label{SI:tab:results_water}
\end{small}
\caption{Overview of the simulated systems. 
    MD = MD integrator or thermostat,
    LF = leap frog, GSD = GROMACS stochastic dynamics, VR = velocity rescale, B = Berendsen, NH = Nosé-Hoover. Left for water and right for organic liquids. The output means the calculation relevant observable with $v_i$ being the velocity of every atom, $P_{xy}$ the system pressure tensor and $E$ the kinetic and total energy of the system. For TIP3P the final box after equilibration is 2.97x2.97x2.97nm³ and for TIP4P 3.02x3.02x3.02 nm³.\\
    $^*$Emulated via massive CSVR thermostat (Bussi-Donadio-Parrinello thermostat with $N_f=1$)
    }
    \label{SI:SI-simulations_water}
\end{table}

\begin{table}
\begin{small}
    \centering
    \begin{tabular}{l|l|| c|c|c}
    \hline
    \multicolumn{5}{c}{organic liquids}\\ \hline
            &               &timestep: 1 fs &timestep: 1 fs &timestep: 1 fs   \\          
            &               &output $P_{xy}$: 5 fs &output $P_{xy}$: 5 fs&output $P_{xy}$: 5 fs\\          
            MD&$\tau$ in ps &pentane &anilin &glycerol 
                                  \\
    \hline\hline
    \multirow{6}{*}{GSD} 
    & 0.01  &20 ns&20 ns&20 ns \\ \cline{2-5}
    & 0.1   &20 ns&20 ns&20 ns \\ \cline{2-5}
    & 1     &20 ns&20 ns&20 ns \\ \cline{2-5}
    & 10    &20 ns&20 ns&20 ns \\ \cline{2-5}
    & 100   &20 ns&20 ns&20 ns \\ \hline

    VR& 1   &20 ns&20 ns&20 ns \\ \hline
    BE & 0.1 &20 ns&20 ns&20 ns \\ \hline    
    NH & 4  &20 ns&20 ns&20 ns \\ \hline
    \end{tabular}
    \label{SI:tab:results_water}
    \caption{Overview of the simulated organic liquid systems. 
    MD = MD integrator or thermostat, GSD = GROMACS stochastic dynamics, VR = velocity rescale, B = Berendsen, NH = Nosé-Hoover. The output means the calculation relevant observable with $P_{xy}$ being the system pressure tensor. For aniline the final box after equilibration is 3.12x3.12x3.12nm³, for or pentane the box is 2.93x2.93x2.93~nm³ and for glycerin the box is 4.19x4.19x4.19~nm³.}
    \label{SI:SI-simulations_organic}
\end{small}
\end{table}

\clearpage
\section{Details for different Liquids}
\label{SI:SI-results}
\subsection{Water}
\label{SI:SI-water}
\begin{table}
    \centering
    \begin{tabular}{l|l|c|c|c|c|c}
        Thermostat& $\tau_T$ in ps &\multicolumn{3}{c|}{D in $10^{-5}$~cm$^2$~s$^{-1}$}  & \multicolumn{2}{c}{$\eta$ in mPas}\\
        & & TIP3P & TIP4P & xTB & TIP3P & TIP4P \\
    \hline
    NVE & 0 & 6.25 $\pm$ 0.0876 & 5.07 $\pm$ 0.0708 & 4.41 $\pm$ 0.173 & 0.297 $\pm$ 0.0484 & 0.534 $\pm$ 0.0802 \\ \hline
    Nose-Hoover & 4 & 6.48 $\pm$ 0.122 & 5.17 $\pm$ 0.169 &  & 0.339 $\pm$ 0.0659 & 0.536 $\pm$ 0.0799 \\ \hline
    Berendsen & 0.1 & 6.52 $\pm$ 0.177 & 5.2 $\pm$ 0.101 &  & 0.319 $\pm$ 0.0389 & 0.503 $\pm$ 0.0732 \\ \hline
    \multirow{6}{*}{Velocity-Rescale} & 0.001 &  &  & 4.44 $\pm$ 0.146 &   &   \\ \cline{2-7}
     & 0.01 & 6.44 $\pm$ 0.0601 & 5.14 $\pm$ 0.0698 & 4.17 $\pm$ 0.165 & 0.337 $\pm$ 0.0419 & 0.501 $\pm$ 0.0778 \\ \cline{2-7}
     & 0.1 & 6.63 $\pm$ 0.187 & 5.19 $\pm$ 0.0408 & 4.39 $\pm$ 0.162 & 0.314 $\pm$ 0.0468 & 0.514 $\pm$ 0.0634 \\ \cline{2-7}
     & 1 & 6.48 $\pm$ 0.109 & 5.16 $\pm$ 0.128 & 4.69 $\pm$ 0.141 & 0.334 $\pm$ 0.0468 & 0.513 $\pm$ 0.0559 \\ \cline{2-7}
     & 10 & 6.41 $\pm$ 0.105 & 5.07 $\pm$ 0.17 &    & 0.333 $\pm$ 0.0522 & 0.526 $\pm$ 0.100 \\ \cline{2-7}
     & 100 & 6.38 $\pm$ 0.23 & 5.05 $\pm$ 0.168 &   & 0.322 $\pm$ 0.0479 & 0.495 $\pm$ 0.0785 \\ \hline
   \multirow{6}{*}{GSD/OBABO} & 0.001 &  &  & 0.09 $\pm$ 0.0024 &  &   \\ \cline{2-7}
     & 0.01 & 1.36 $\pm$ 0.00971 & 1.38 $\pm$ 0.0121 & 0.6 $\pm$ 0.018 & 2.62 $\pm$ 0.478 & 3.81 $\pm$ 0.62 \\ \cline{2-7}
     & 0.1 & 3.37 $\pm$ 0.0169 & 3.02 $\pm$ 0.0637 & 2.71 $\pm$ 0.123 & 0.583 $\pm$ 0.109 & 0.865 $\pm$ 0.147 \\ \cline{2-7}
     & 1 & 5.7 $\pm$ 0.0541 & 4.71 $\pm$ 0.0539 & 3.98 $\pm$ 0.123 & 0.369 $\pm$ 0.0633 & 0.571 $\pm$ 0.130 \\ \cline{2-7}
     & 10 & 6.35 $\pm$ 0.131 & 5.0 $\pm$ 0.0392 &  & 0.355 $\pm$ 0.0667 & 0.512 $\pm$ 0.103 \\ \cline{2-7}
     & 100 & 6.41 $\pm$ 0.216 & 4.76 $\pm$ 0.156 &  & 0.358 $\pm$ 0.0535 & 0.516 $\pm$ 0.102 \\
        \end{tabular}
    
    \caption{Overview of the results of Diffusion constant D and viscosity $\eta$ for different thermostats and different coupling times $\tau_T$. The shown error is the standard deviation calculated from block averaging.}
    \label{SI:SItab:results_water}
\end{table}

\begin{figure}[H]
    \centering \includegraphics[width=\columnwidth]{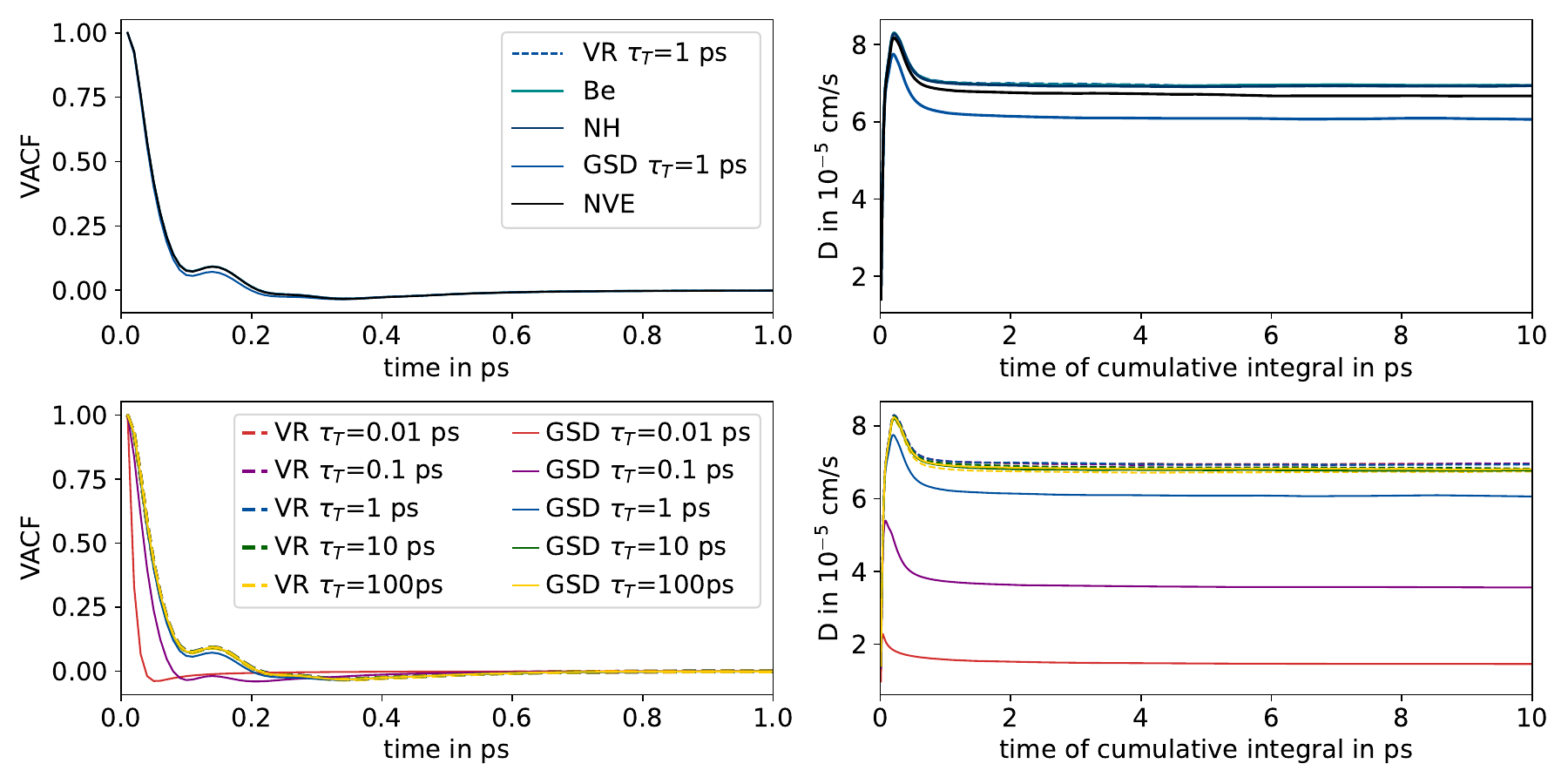}
    \caption{VACF and diffusion constant plots for TIP3P}
    \label{SI:fig:TIP3P_VACF}
\end{figure}

\begin{figure}[H]
    \centering \includegraphics[width=\columnwidth]{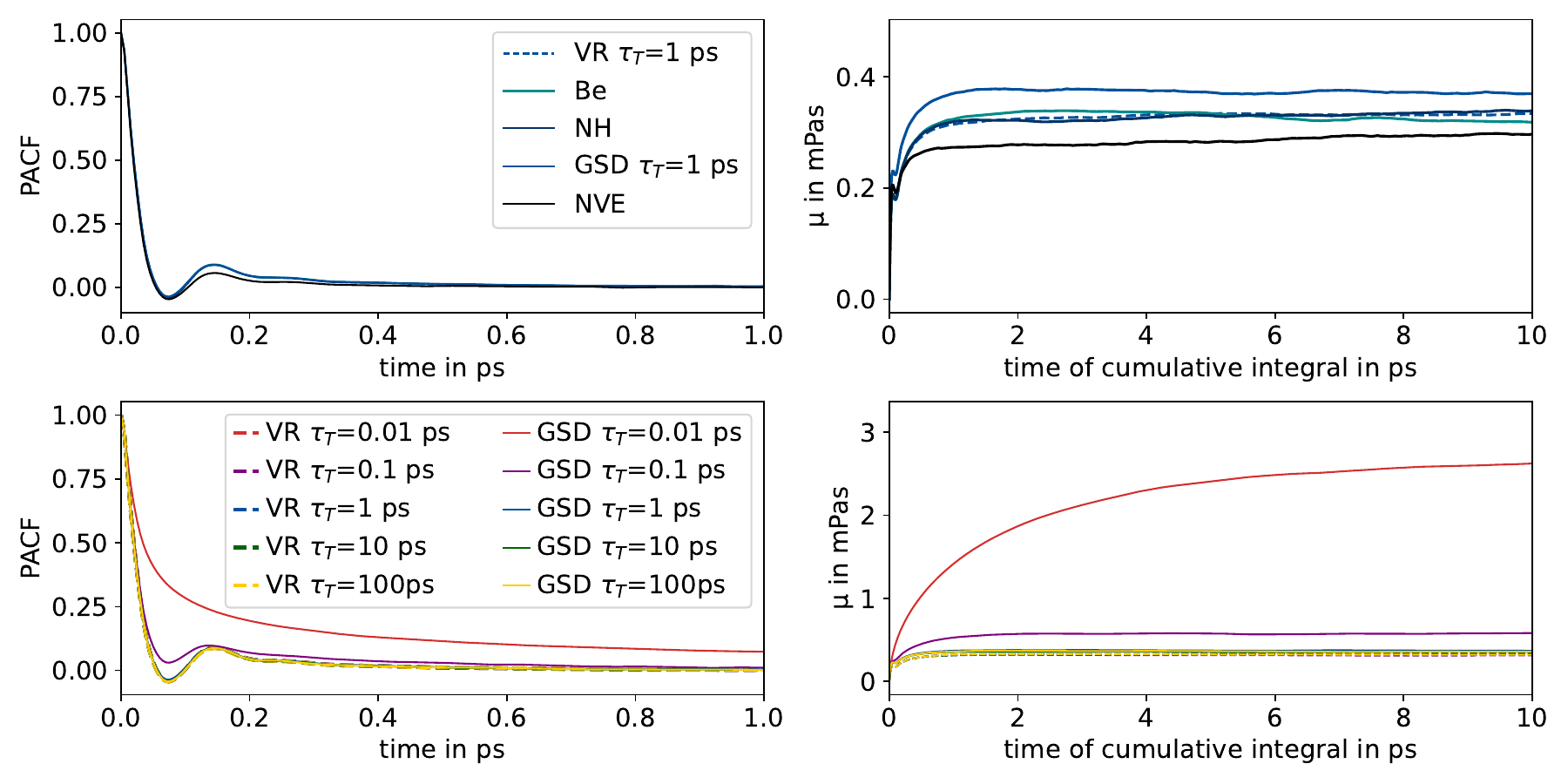}
    \caption{PACF and viscosity plots for TIP3P}
    \label{SI:fig:TIP3P_PACF}
\end{figure}

\begin{figure}[H]
    \centering \includegraphics[width=\columnwidth]{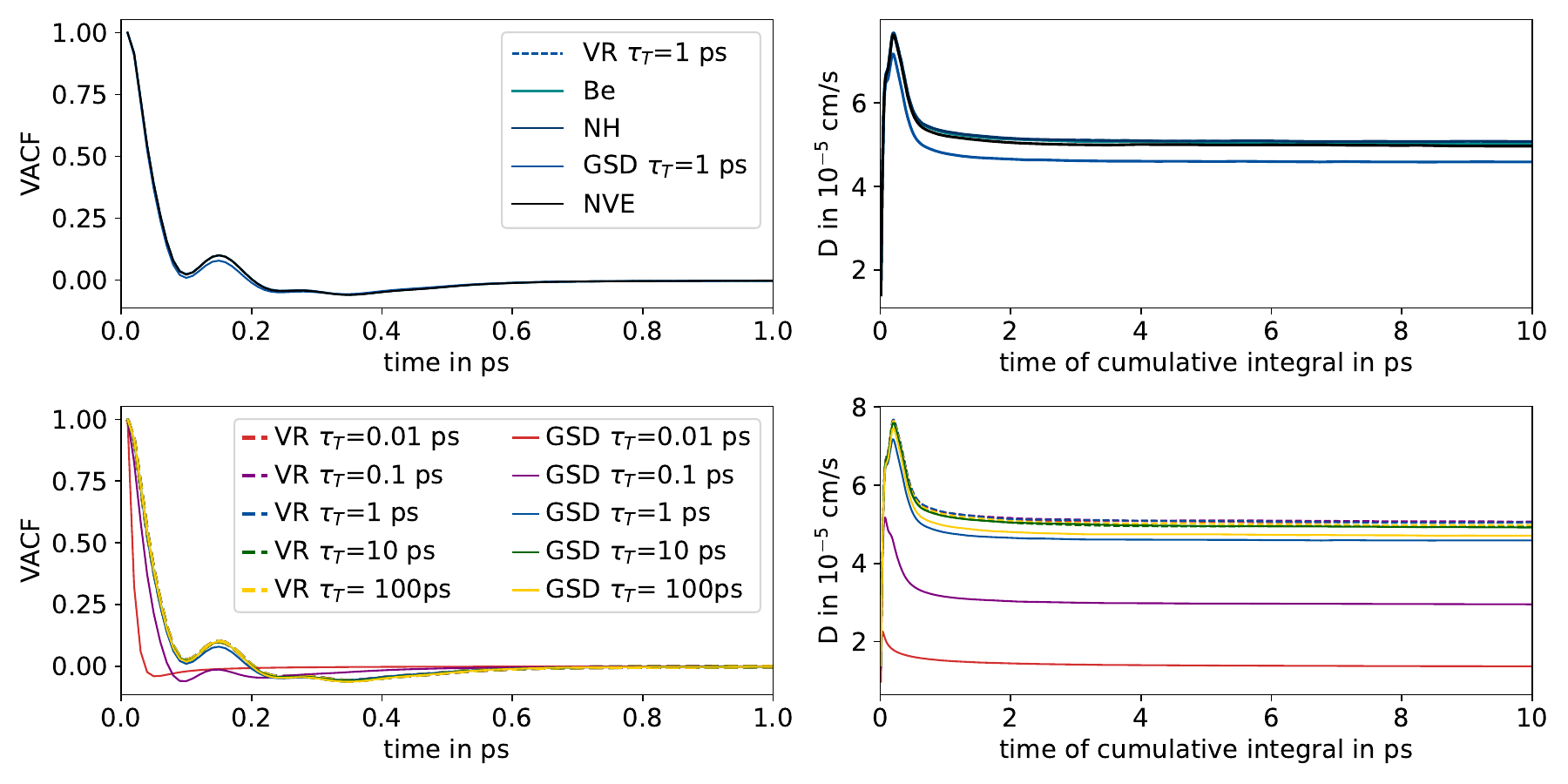}
    \caption{VACF and diffusion constant plots for TIP4P}
    \label{SI:fig:TIP4P_VACF}
\end{figure}

\begin{figure}[H]
    \centering \includegraphics[width=\columnwidth]{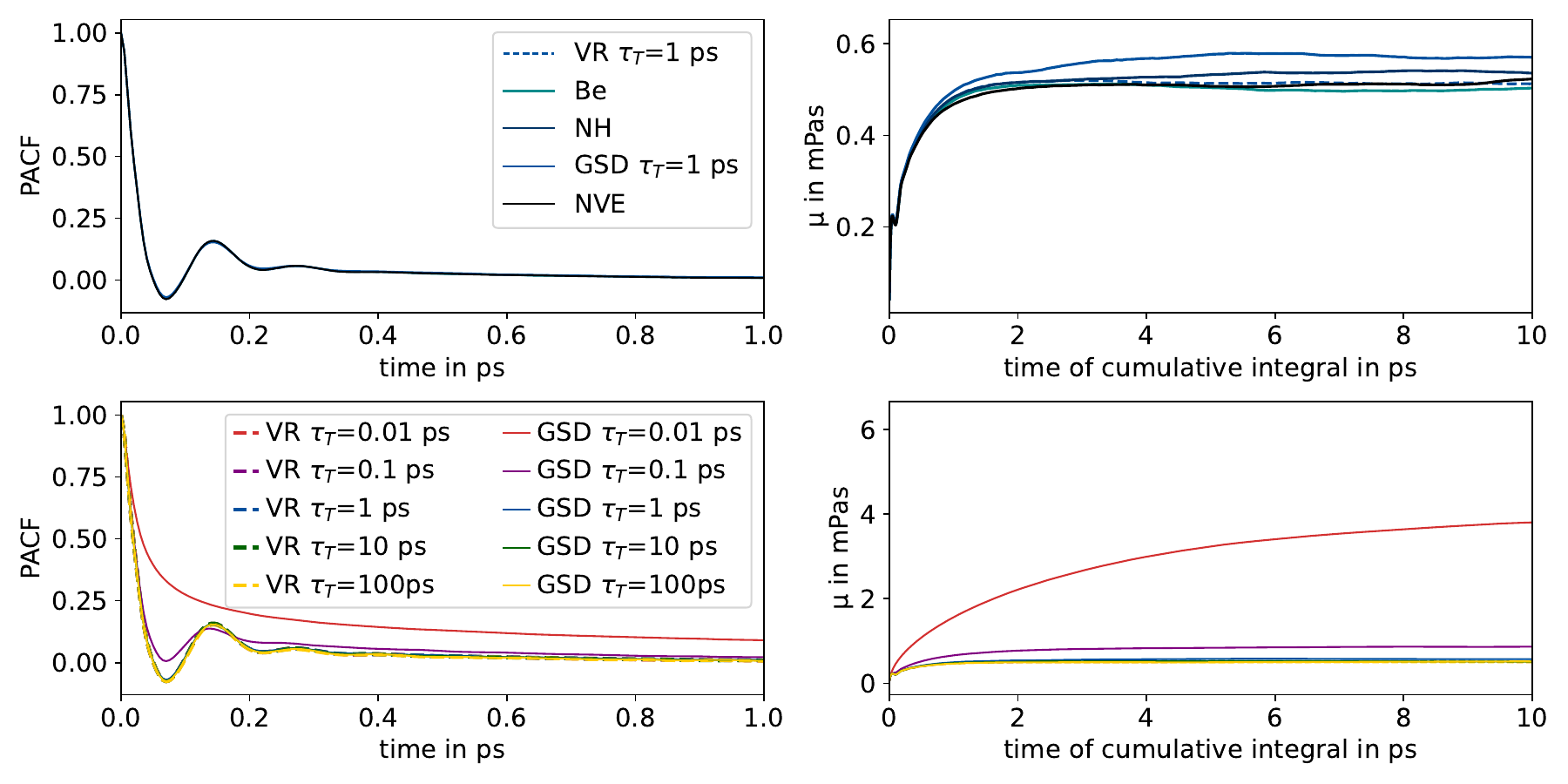}
    \caption{PACF and viscosity plots for TIP4P}
    \label{SI:fig:TIP4P_PACF}
\end{figure}

\begin{figure}[H]
    \centering
    \includegraphics[scale=0.45]{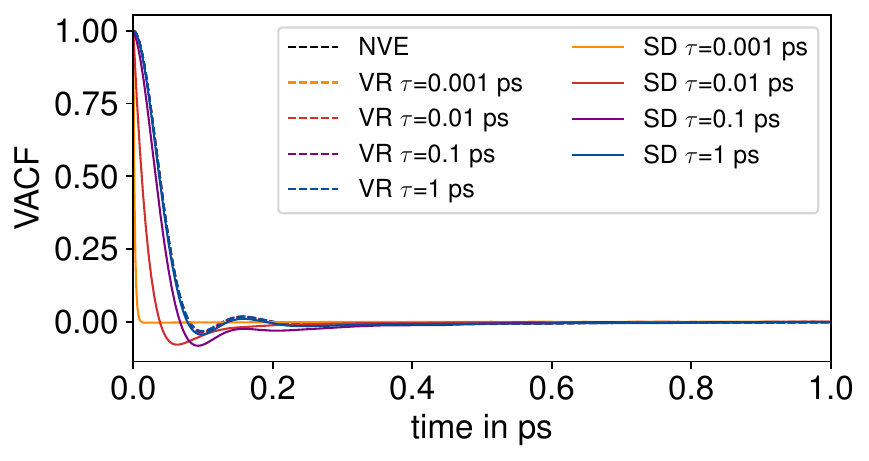}
    \includegraphics[scale=0.45]{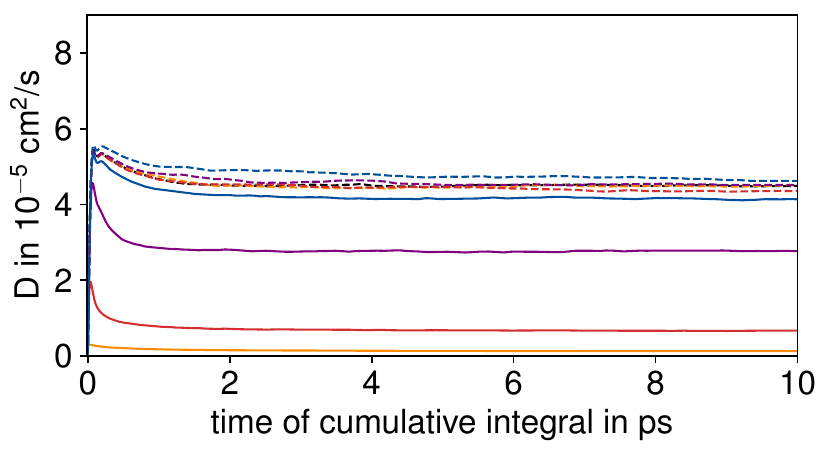}
    \caption{Plots for xTB (Molecular VACF)}
    \label{SI:fig:xTB}
\end{figure}

\newpage
\subsection{Organic Liquids}
\label{SI:SI-Organic}
\begin{table}[h]
    \centering
    \begin{tabular}{l|l|c|c|c}
        Thermostat& $\tau_T$ in ps & \multicolumn{3}{c}{$\eta$ in mPas}\\
        & & Pentane & Anilin & Glycerol \\
    \hline
    Experimental\cite{viscosities} & & 0.24 & 4.40 & 1490\\ \hline 
    Nose-Hoover & 4 & 0.107 $\pm$ 0.0249 & 1.92 $\pm$ 0.849 &  1710 $\pm$ 1680\\ \hline
    Berendsen & 0.1 & 0.105 $\pm$ 0.0266 & 1.93 $\pm$ 0.636 &  1880 $\pm$ 1440\\ \hline
    Velocity-Rescale & 1 &  0.116 $\pm$ 0.0256 & 2.0 $\pm$ 0.886 & 1600  $\pm$ 1450\\ \hline
   \multirow{5}{*}{Stochastic Dynamics} & 0.001 & 1.69  $\pm$ 0.665 & 23.4 $\pm$ 17.4 &  2220 $\pm$ 905\\ \cline{2-5}
     & 0.1 &  0.229 $\pm$ 0.0701 & 6.1 $\pm$ 3.78 &  1170 $\pm$905\\ \cline{2-5}
     & 1 &  0.139 $\pm$ 0.0476 & 2.2 $\pm$ 1.42 &  992 $\pm$ 713\\ \cline{2-5}
     & 10 &  0.114 $\pm$ 0.0362 & 1.75 $\pm$ 0.761 & 1910  $\pm$ 999\\ \cline{2-5}
     & 100 & 0.101 $\pm$ 0.0265 & 2.74 $\pm$ 1.71 &  2180 $\pm$ 1780\\
        \end{tabular}
    
    \caption{Overview of the results viscosity $\eta$ for different thermostats and different coupling times $\tau$ of pentane, anillin and glycerol. The shown error is the standard deviation calculated from block averaging.}
    \label{SI:SItab:viscosities}
\end{table}

\begin{figure}[H]
    \centering
    \includegraphics[width=\columnwidth]{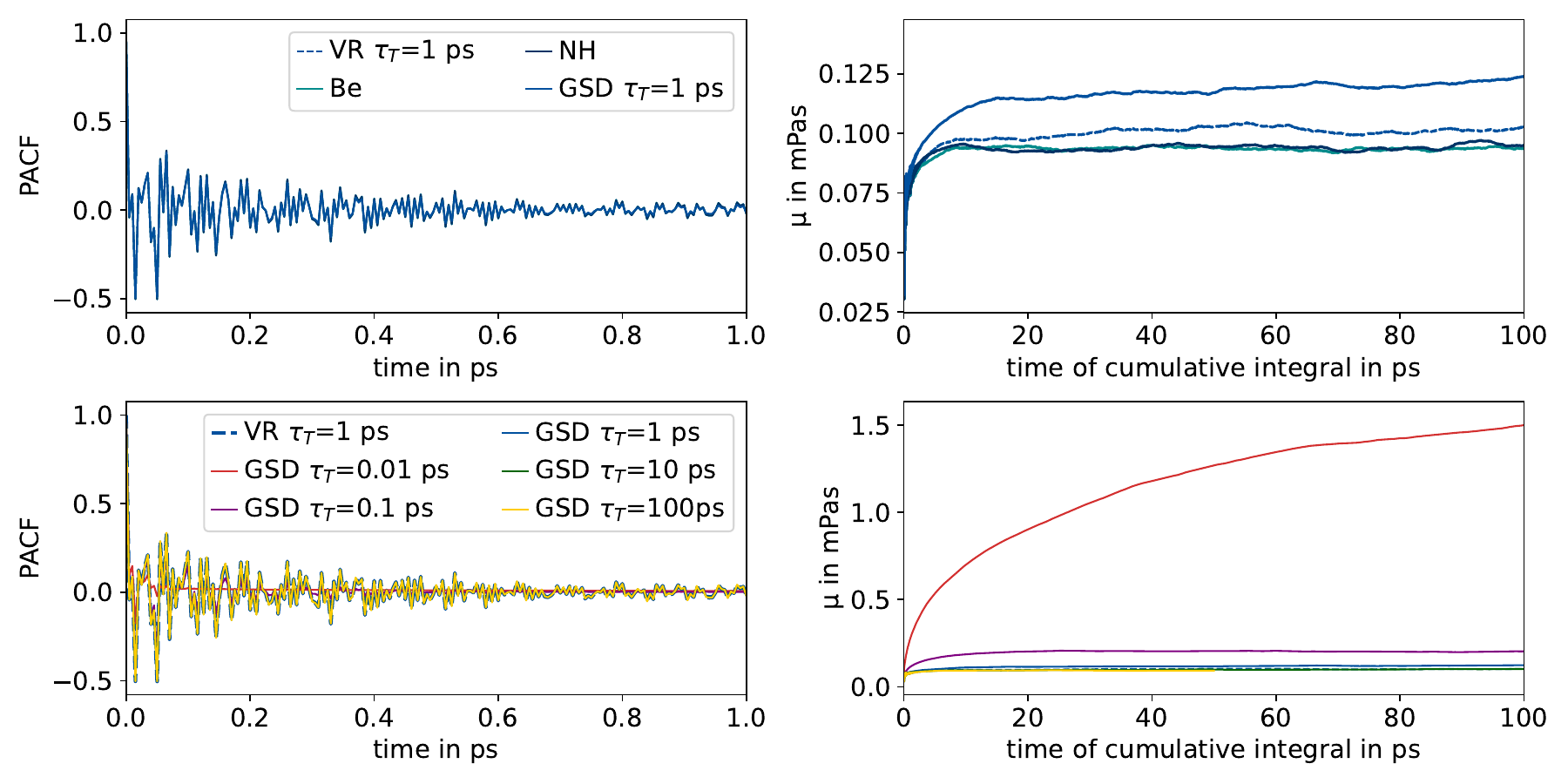}
    \caption{Plots for Pentane}
    \label{SI:SIfig:Pentane}
\end{figure}
\begin{figure}[H]
    \centering
    \includegraphics[width=0.4\linewidth]{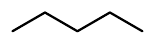}
    \caption{Structure of pentane}
    \label{SI:SIfig:Pentan_structure}
\end{figure}

\begin{figure}[H]
    \centering
\includegraphics[width=0.85\columnwidth]{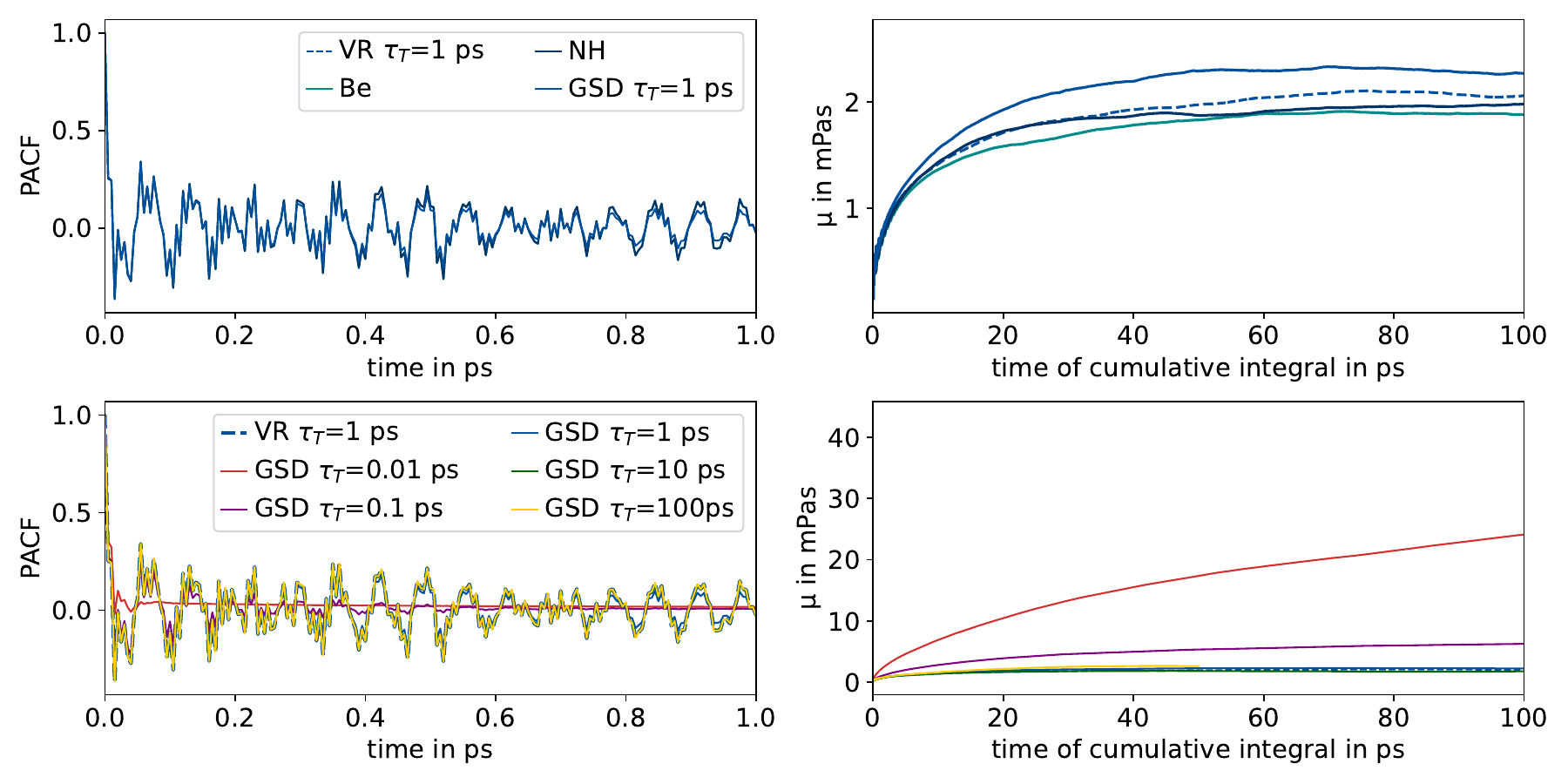}
    \caption{Plots for Anilin}
    \label{SI:SIfig:Anilin}
\end{figure}

\begin{figure}[H]
    \centering
    \begin{minipage}{.5\textwidth}
        \centering
        \includegraphics[width=0.6\linewidth]{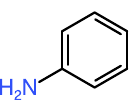}
        \caption{Structure of anilin}
        \label{SI:SIfig:Anilin_structure}
    \end{minipage}%
    \begin{minipage}{0.5\textwidth}
        \centering
        \includegraphics[height=2.9cm]{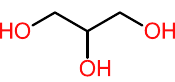}
        \caption{Structure of glycerol}
        \label{SI:SIfig:Glycerol_structure}
    \end{minipage}
\end{figure}

\begin{figure}[H]
    \centering
    \includegraphics[width=0.85\columnwidth]{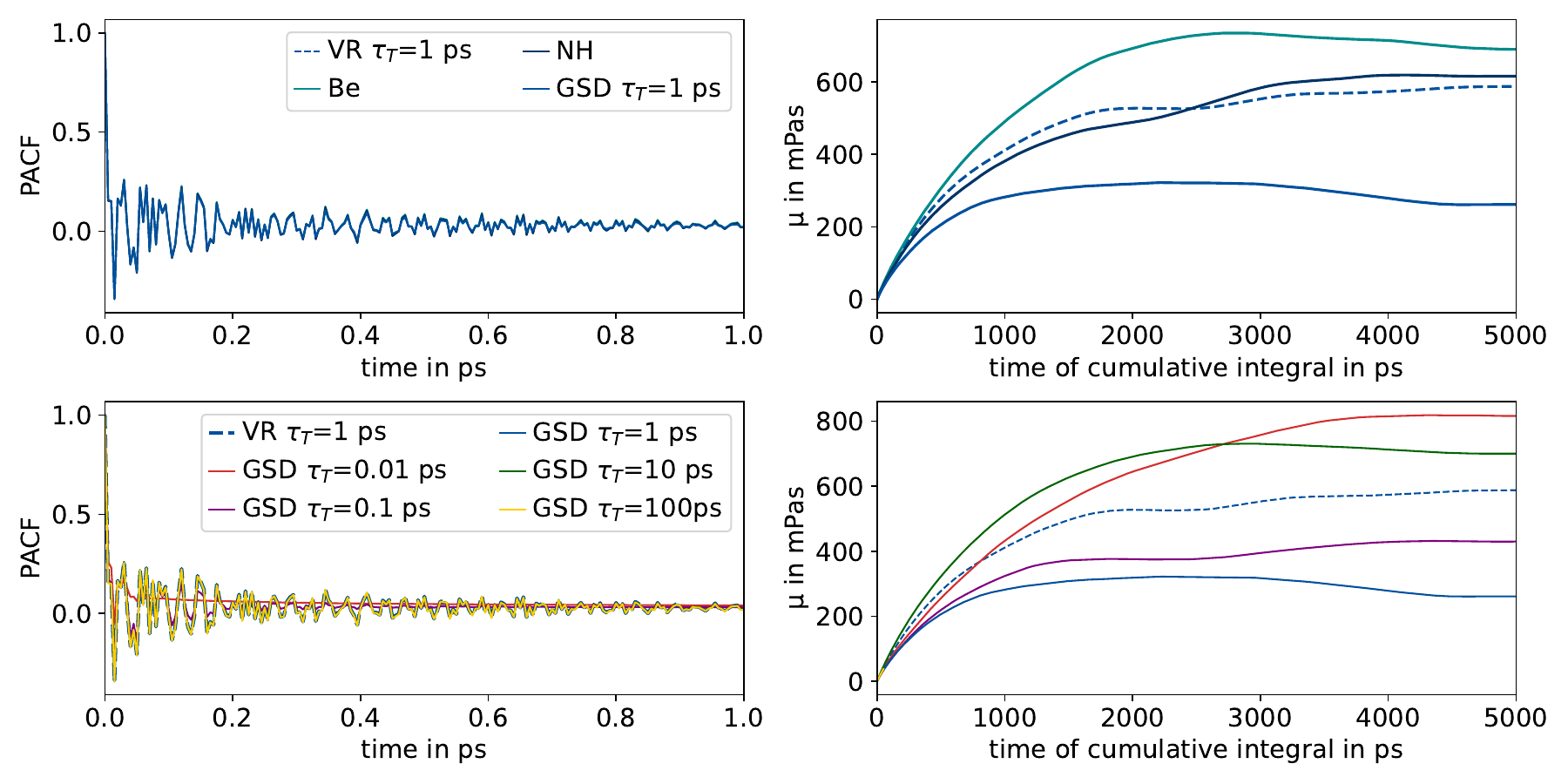}
    \caption{Plots for glycerol}
    \label{SI:SIfig:Glycerine}
\end{figure}
\newpage
\section{Additional theory}
\label{SI:SI:Theory}
\subsection{Transport phenomena}
Consider a scalar physical quantity $Q(t)$ that is distributed in the three-dimensional space and flows through a surface $A$.
The total current 
\begin{eqnarray}
    I(A) &=& \frac{dQ}{dt}
\end{eqnarray}
represents the total amount of $Q$ that passes through $A$ per unit time.
The flux, or equivalently current density, 
\begin{eqnarray}
    \mathbf{J} = \frac{dQ}{dAdt}
\end{eqnarray}
represents the amount of $Q$ that flows per unit time and unit area and the direction of this flow.
Thus, $\mathbf{J} \in \mathbb{R}^3$. 
The total current can be calculated from the flux by integrating over the surface $A$
\begin{eqnarray}
    I(A) = \iint_{A} \mathbf{J} \cdot \mathrm{d}\mathbf{A} = \iint_{A} \mathbf{J} \cdot  \mathbf{n}\,  \mathrm{d}A\, ,
\end{eqnarray}
where $\mathbf{n}$ is the normal vector on the surface $A$, and 
$\mathbf{J} \cdot  \mathbf{n}$ denotes the Euklidean scalar product between $\mathbf{J}$ and $\mathbf{n}$.
Since the flux depends on the position at which the flow through $A$ is measured, the flux is a three-dimensional vector field $\mathbf{J}: \mathbb{R}^3 \rightarrow \mathbb{R}^3$.
In linear response theory, the $\mathbf{J}$ is linearly related to a second vector field $\mathbf{F}: \mathbb{R}^3 \rightarrow \mathbb{R}^3$, which causes the flux: 
\begin{eqnarray}
    \mathbf{J} = \kappa \, \mathbf{F} \, ,
\end{eqnarray}
where $\kappa$ is the transport coefficient.
Switching to a microscopic picture, the continuous quantity $Q(t)$ can be related to a microscopic quantity $a(t)$ that is defined for each particle of the system. 
The Green-Kubo relations state that the transport coefficient can be calculated from the time auto-correlation function of the time-derivative of $a(t)$
\begin{eqnarray}
    \kappa = \frac{1}{k_BTV}\int_0^{\infty} \langle \dot{a}(t)\dot{a}(0) \rangle \, \mathrm{d}t.
\end{eqnarray}
Since time auto-correlation functions are easily evaluated from MD simulations, this gives an easy way to analyze transport processes, which are macroscopic properties, from simulations of systems with relatively few particles.

\subsection{Onsager regression hypothesis and correlation functions}
The Onsager regression hypothesis (1931)\cite{OnsagerI},\cite{OnsagerII}, which is one of the foundations of statistical thermodynamics, is a consequence of the historically later defined fluctuation-dissipation theorem (1966) FDT by Kubo\cite{Kubo1966}. The FDT links the dissipation of energy through friction with the statistical thermodynamic fluctuations. From this follows Onsagers regression theory, which states, that the relaxation of statistical fluctuations is indistinguishable from the relaxation of small disturbances from the equilibrium. 
This allows equilibrium MD simulations an easy access to an observable like the diffusion constant, which is in Fick's law\cite{Fick} a non equilibrium process. The usage of systems in equilibrium allows for the sampling of a much smaller phase space. The relaxation of fluctuations can mathematically described with correlation functions.

The most common use of ACF in MD simulations is through the use of Green-Kubo relations, which link a macroscopic observable $A$ to the integral over an auto-correlation function of the corresponding microscopic fluctuation $a$. 
\begin{equation}
     A=F\int_0^t \langle a(0)a(0+\tau_{{\mathrm{lag}}})\rangle \mathrm{d}\tau_{\mathrm{lag}}
\end{equation}
$F$ is a proportionality factor depending in the system. Since auto-correlations are time invariant it is custom to start at $t=0$. While we do not want to co into too much detail of their derivation, we want to at least share the steps and ideas behind it.

\subsection{Diffusion}
The derivation of the Green-Kubo relation for the diffusion constant D is a relative straight forward solving of a partial differential equation as its main problem.\cite{Helfand}
We first start with the diffusion by using Fick's 2nd law\cite{Fick}for one dimension, which links the change of a concentration over time with the change of a concentration over space through the diffusion constant D. For our case the concentration c is unit less and is 1 at time $t=0$ and the origin position $x_0$.
\begin{equation}
    \frac{\partial c}{\partial t}=D\frac{\partial^2 c}{\partial x^2}
\end{equation}
Besides our initial concentration the second boundary condition is that in an infinite system, the "wall" can not be reached in finite time and the concentration is 0. For this differential equation the solution is known in the form of a Gaussian function:
\begin{equation}
    c = \frac{1}{\sqrt{4\pi Dt}} e^{-\frac{(x-x_0)^2}{4Dt}}
\end{equation}
Since this is a probability distribution function we can make use of its variance to obtain the diffusion constant.
\begin{equation}
    \langle (x-x_0)^2 \rangle =\int_{-\infty}^{\infty}  (x-x_0)^2
    \cdot c(x,t)  \mathrm{d}x =2Dt
\end{equation}
The variance is nothing else but the ensemble average of the squared displacement or mean squared displacement. With $\Delta x= (x-x_0)$ and for a single particle i this results in the known mean squared displacement MSD equation in the limit of long time spans:
\begin{equation}
    D_i=\frac{1}{2t} \langle (\Delta x_{i})^2 \rangle 
    \label{SI:eqn:MSD}
\end{equation}
A position difference can also be expressed as an integral over a velocity $v$.
\begin{equation}
    \Delta x_i=\int_{0}^{t} v_i  \mathrm{d}\tau
\end{equation}
Through substitution and the clever usage of several general properties of auto-correlation functions we can rewrite Eqn. \ref{SI:eqn:MSD}
\begin{equation}
    \langle (\Delta x_{i})^2 \rangle =  2t\int_{0}^{\infty} \langle v(t)v(0) \rangle  \mathrm{d}t 
\end{equation}
Combined with our previous result we reach the Green-Kubo relation for the self diffusion constant $D$. 
\begin{equation}
    D=\int_{0}^{\infty} \langle v(t)v(0) \rangle  \mathrm{d}t
\end{equation}

\subsection{Shear viscosity}
For the shear viscosity $\eta$ we link the momentum fluctuation $p$ to the density fluctuation $\rho$ with the help of the particle current $j$.\cite{Hansen2013}
For an one component fluid witch follows the laws of conservation the number density $\rho$ and momentum density $\bm{p}$ can be expressed trough "heat equations". For a system with a mean velocity of zero the differential equations are:
\begin{equation}
m \frac{\partial}{\partial t} \rho(\bm{r},t)+\nabla \cdot \bm{p}(\bm{r},t) = 0
    \label{SI:eqn:density_dissipation}
\end{equation}
\begin{equation}
\frac{\partial}{\partial t} \bm{p} (\bm{r},t)+\nabla \cdot \Pi(\bm{r},t) = 0
    \label{SI:eqn:momentum_dissipation}
\end{equation}
For the shear viscosity we only need the stress tensor $\Pi$ where $\alpha \neq \beta$ e.g. the off diagonal elements which results in a simple $\Pi$:
\begin{equation}  
\Pi^{\alpha\beta}=\eta \left(  \frac{\partial u_{\alpha}(\bm{r},t)}{\partial r_{\beta}} +\frac{\partial u_{\beta}(\bm{r},t)}{\partial r_{\alpha}} \right)
\label{SI:eqn_stress_tensor}
\end{equation}
We furthermore assume the local deviation for the number density is small, which allows us to express the momentum density as the particle current:
\begin{equation}
    \bm{p}(\bm{r},t) \approx m\rho \bm{u}(\bm{r},t) \equiv m\bm{j}(\bm{r},t)
    \label{SI:eqn:current}
\end{equation}
After substitution we reach:
\begin{equation}
    \frac{\partial}{\partial t} \bm{j} (\bm{r},t)-\frac{1}{m}\nabla \cdot \frac{\eta}{\rho} \left(  \frac{\partial j_{\alpha}(\bm{r},t)}{\partial r_{\beta}} +\frac{\partial j_{\beta}(\bm{r},t)}{\partial r_{\alpha}} \right)  = 0 
\end{equation}
Since we only look at the x direction we can simplify to:
\begin{equation}
    \frac{\partial}{\partial t} \bm{j}^x (\bm{r},t)-\frac{\eta}{\rho m} \nabla^2 \bm{j}^x(\bm{r},t)  = 0 
\end{equation}
To solve this partial differential equation we use the Laplace and Fourier transformation, which transforms the problem into an ordinary differential equation, which we can solve. In the end we link the microscopic stress tensor to the macroscopic pressure tensor and obtain the Green-Kubo relation for shear viscosity:
\begin{equation}
    \eta= \frac{1}{Vk_BT} \int_0^{\infty} \langle P^{xz}(t){P^{xz}(0)}\rangle\mathrm{d}t
    \label{SI:eqn:GK_viscosity}
\end{equation}
$V$ is the volume of the system, $T$ the temperature, $k_B$ the Boltzmann constant and $P^{xz}$ the off diagonal elements of the system pressure tensor

\subsection{MSM}
\begin{figure}[H]
    \centering
    \includegraphics[width=0.9\columnwidth]{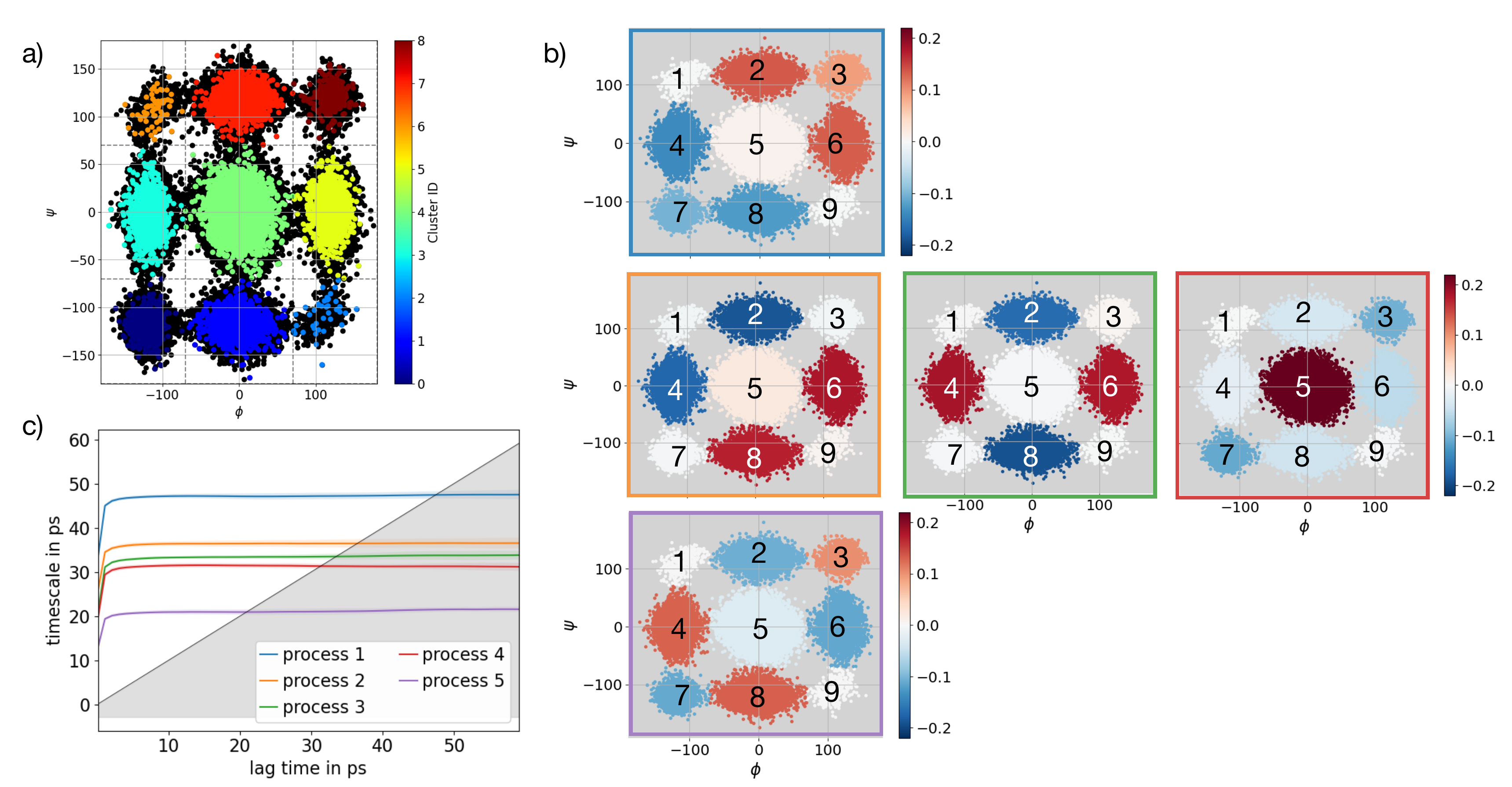}
    \caption{
    a) MSM grid with 9 mircostates in ($\phi$,$\psi$)-angle space.
    b) The MSM eigenvectores at $20.1$~ps for the process 1-5. 
    c) MSM implied timescales with standard deviation as function of lag times in the range of $0.1-59.1$~ps for the first five slowest processes.
    }
    \label{SI:SI_fig:MSM}
\end{figure}

The dynamical properties of the first five processes are represented for a MSM of 9 states at lag time $\tau=20.01$~ps and coupling parameter $\tau_T=1$~ps. 
Fig.~\ref{SI:SI_fig:MSM} is a representative example of the dynamics of all analyzed cases with different coupling parameter.
The discretization is shown in Fig.~\ref{SI:SI_fig:MSM}.a). 
The dynamics is split into a slowest process ($\lambda_1$, $\mathbf{l}_1$ blue), three similarly fast processes ($\lambda_2$, $\mathbf{l}_2$ orange; $\lambda_3$, $\mathbf{l}_3$ green; $\lambda_4$, $\mathbf{l}_4$ red), and further separate processes ($\lambda_5$, $\mathbf{l}_5$ purple); see Fig.~\ref{SI:SI_fig:MSM}.c).
The slowest process is characterised by transitions along the diagonal running through clusters 1,5 and 9. 
Processes 2 and 3 are maily jumps between cluter 2,4,6 and 8.
The next slower processes shows the exchange between the most populated cluster 5 and the surrounding regions.
The slowest process shown here is the exchange of the least popular clusters with 2,4,6 and 8.


\end{document}